\documentstyle{espart}

\input epsf

\begin{document}
\begin{frontmatter}

\hfill TAN-FNT-96/003

\hfill hep-th/9612129

\title{Covariant Quantization of the Skyrmion}

\author{M. Kruczenski \thanksref{email}}
\address{Departamento de F\'{\i}sica, Comisi\'on Nacional de Energ\'{\i}a
At\'omica, Av.\ del\ Libertador 8250, 1429 Buenos Aires, Argentina.}
\and
\author{L.E. Oxman\thanksref{CBPF}}
\address{Instituto de Fisica, Universidade Federal do Rio de Janeiro,
C.P. 68528, Rio de Janeiro, Brasil.}
\thanks[email]{E-mail: martink@tandar.cnea.edu.ar}
\thanks[CBPF]{Partially supported by CLAF/CNPq, Brasil.} 

\begin{abstract}
We obtain the internal degrees of freedom of the skyrmion (spin and
isospin) within a manifestly Lorentz covariant quantization framework
based on defining Green functions for skyrmions and then, the S-matrix
via LSZ reduction. Our method follows Fr\"ohlich and Marchetti's
definition of Euclidean soliton Green functions, supplemented with a
careful treatment of the boundary conditions around the singularities.
The covariant two-point function obtained propagates a tower of spin
equal to isospin particles. Our treatment contains the usual
method of collective coordinates, as a non-relativistic limit and,
because of the new topology introduced, it leads, in a natural way, to 
the inequivalent (boson/fermion) quantizations of the $\mathrm{SU}(2)$ 
skyrmion.

\end{abstract}
\end{frontmatter}

%\pacs{11.10.Lm}

\section{Introduction}

Since the work of Witten~\cite{Wi79,Wi83}, the Skyrme model~\cite{Sk61}
has received renewed attention as the low energy limit of QCD, as it is
supposed to describe QCD in the large number of colors ($N_c$) limit.
Recently,  B. Moussallam~\cite{M93} gave further support to the soliton
picture showing that the low energy expansion of chiral perturbation
theory~\cite{GL84} supplemented with a large $N_c$ analysis is
compatible with a stable soliton.

The large $N_c$ limit can be thought of as a semiclassical expansion,
as $N_c$ appears in the combination $\hbar/N_c$~\cite{W79}. One starts
from a minimum of the potential which is the classical soliton profile.
The soliton solution has six zero modes corresponding to translations
and rotations (or isorotations). Quantization of
 these zero modes gives rise to linear momentum, spin and isospin. An
analysis of the wave functions~\cite{Wi79} reveals the existence of a
tower of particles with spin equal to isospin.

Also, as the mass of the skyrmion scales as $N_c$, in the large $N_c$
limit a non-relativistic treatment of the skyrmion is justified.
Although this is correct in principle, it is not always desirable to
break Lorentz invariance as one looses the possibility of applying
standard knowledge from quantum field theory. For example, in
reference~\cite{DHM94}, the equivalence to a relativistic field theory
was advocated as an important point to solve the so called Yukawa
problem. Therefore it is useful to introduce a manifestly Lorentz
covariant quantization, defining Green functions for skyrmion fields.
We believe that it is conceptually better to have a definition for
these Green functions and, a posteriori, to take the non-relativistic
limit.

Fr\"ohlich and Marchetti~\cite{FM90} introduced Euclidean Green
functions for skyrmions following previous developments on soliton 
quantization~\cite{FM87,MS80}. In reference~\cite{tH78} similar ideas
were considered in the context of gauge theories.

In the present paper we will consider the covariant approach of
ref.~\cite{FM90} but, in order to obtain the skyrmion internal degrees
of freedom (spin and isospin), we will prescribe the boundary
conditions which the field satisfies around the singularities. This
treatment will allow us to relate the covariant quantization approach
to more traditional ones for skyrmion quantization\footnote{ For a
review see~\cite{ZB86}}.  Also, this covariant quantization is
a natural candidate for the skyrmion quantum field theory underlying
the work of N. Dorey, J. Hughes and M.P. Mattis~\cite{DHM94}.

In section~\ref{sec:Gf} the Green functions defined in~\cite{FM90} are
reviewed. In section~\ref{sec:bc} the boundary conditions
 are discussed, showing that they bring in new topology. In
section~\ref{sec:sotp} the propagator is shown to transform correctly
under Lorentz and isospin transformations. Further arguments are given
 in section~\ref{sec:Sp} to show that the propagator describes a tower
of spin equal to isospin particles. In section~\ref{sec:Yc} we give the
LSZ rules to construct the S-matrix and discuss the relation of our
calculation to the computation of ref.~\cite{DHM94}. 
In section~\ref{sec:C} our conclusions are given.
Finally, in an appendix, we summarize the definition of arbitrary spin
propagators due to Weinberg~\cite{W64}.

\section{Green functions } \label{sec:Gf}

The Skyrme model is defined by the following Lagrangian
\begin{equation}
{\cal L} = \frac{f_\pi^2}{4} {\rm Tr} \partial_{\mu}U^{\dagger}
\partial^{\mu}U + \frac{1}{32 e^2} {\rm Tr}\left(
\left[ U^{\dagger}\partial_{\mu}U, U^{\dagger}\partial_{\nu}U 
\right]^2 \right) ,
  \label{eq:lag}
\end{equation}
where $U$ is a matrix belonging to $\mathrm{SU(2)}$. The lagrangian is
invariant under global $\mathrm{SU(2)}_L\times \mathrm{SU(2)}_R$ chiral
symmetry transformations $U\rightarrow g_L U g_R^\dagger$, $g_L,g_R\in
\mathrm{SU(2)}$. This symmetry is spontaneously broken. If one chooses
the vacuum expectation value of $U$ to be $\langle U\rangle=1$, the
surviving symmetry is the diagonal subgroup $\mathrm{SU(2)}_V$
$(g_L=g_R)$.  A convenient parametrization of $U$ is
\begin{equation}
U=\exp(\mathrm{i} \phi_a \tau^a) ,
\label{pions}
\end{equation}
where $\tau^a$ are the Pauli matrices, and the fields $\phi_a$ 
describe the pions which have zero vacuum expectation value
($\langle U\rangle=1 \Rightarrow \langle\phi_a\rangle =0$).
One can also consider a pion mass term and higher derivative terms such
as $-\epsilon_6 B^\mu B^\mu$, where $\epsilon_6$ is a constant
and $B^\mu$ is defined in eq. (\ref{eq:B}) below. From now on, we will 
work in Euclidean space and will use upper indices ($\mu=1,...,4$)
to denote space-time components. In the following, $4$-dimensional Euclidean
rotations will be called Lorentz transformations for the sake of brevity and
to distinguish them from $3$-dimensional rotations. However this is an abuse
of language and the reader should keep in mind that we are working in 
Euclidean space-time. 

Static configurations with finite energy must satisfy
$U\rightarrow 1$ at infinity, this amounts to compactify space
to $S^3$. Thus, a static configuration defines a map 
$S^3 \rightarrow \mathrm{SU(2)}$, that is, an element of
$\Pi_3(S^3)=Z$. The baryonic number $B$, defined as the integral
of the topological current
\begin{eqnarray}
B^{\mu} &=& \frac{\epsilon^{\mu\nu\alpha\beta}}{24\pi^2}
{\rm Tr}\left[\left(U^\dagger\partial^{\nu}U\right)
\left(U^\dagger\partial^{\alpha}U\right)
\left(U^\dagger\partial^{\beta}U\right)\right] , \nonumber\\
B &=& \int d^3x\ B^4 ,
 \label{eq:B}
\end{eqnarray}
is the winding number associated with $\Pi_3(S^3)$ and 
therefore is conserved.
Configurations with $B=1$ describe nucleon states. The energy of the $B=1$
static configuration that minimizes the action is concentrated
in space; then, this configuration can be approximately treated  as a rigid 
body rotating in space and also in internal space. Spin and isospin
arise quantifying these collective degrees of freedom~\cite{Wi79}.

If there are quantum states associated with the classical soliton 
then it should be possible to define Green functions such as
\begin{equation}
G^{n_1\ldots n_p, a_1\ldots a_r}(x_1\ldots x_p,y_1\ldots y_r) 
= \langle T\left(  
\psi^{n_1}(x_1)\ldots \psi^{n_p}(x_p) 
\phi_{a_1}(y_1)\ldots \phi_{a_r}(y_r)\right)\rangle ,
 \label{eq:Gf}
\end{equation}
where $\psi^{n_i}(x_i)$ (formally) represents an operator destroying
a soliton of winding number $n_i$ at $x_i$ or creating an 
antisoliton $-n_i$ at $x_i$. 

Following~\cite{FM90} we introduce a path integral over fields defined
on space-time compactified to $S^4$ and with the points 
$x_1\ldots x_p$ removed
($M_{1\ldots p}=(R^4\cup \{\infty\})\backslash\{x_1,\ldots,x_p\}$). 
 The manifold $M_{1\ldots p}$ is contractible to 
a bundle of $p-1$ spheres $S^3$ with one common point, implying that maps
from $M_{1\ldots p}$ to $\mathrm{SU(2)}$ are classified by $p-1$ independent 
integers. These integers can be written in terms of $p$ integers $n_i$, 
satisfying $n_1+\cdots+n_p=0$, defined by
\begin{equation}
n_i = \int_{S_i} \mathrm{d}S\  n^{\mu} B^{\mu}\,\, , i=1\ldots p ,
\end{equation}
where $S_i$ is a three sphere enclosing $x_i$ which does not enclose any 
other point $x_j$ ($j\neq i$), $n^\mu$ is the normal to $S_i$ and 
$\mathrm{d}S$ is the surface element on $S_i$. 
Due to the fact that $U$ is constant at infinity, we have
$n_1+\cdots+n_p=0$.

The skyrmion Green functions in eq. (\ref{eq:Gf}) were defined by
Fr\"ohlich and Marchetti in ref. \cite{FM90}. Including the pion field
definition (\ref{pions}), the mixed pion-skyrmion Green functions
read
\begin{equation}
G^{n_1\ldots n_p, a_1\ldots a_r}(x_1,\ldots,x_p,y_1,\ldots y_r) =
\int {\cal D}[U] \phi_{a_1}(y_1)\ldots \phi_{a_r}(y_r)
\exp(-\int_M {\cal L}[U]) ,
\label{eq:prop}
\end{equation}
(see \cite{FM90} for an interpretation in terms of open line defects).
$U=\exp(\mathrm{i} \phi_a \tau^a)$ takes the value $U=1$ at space-time
infinity and has topological number $n_i$ around $x_i$ ($i=1,\ldots
,p$).  A rigorous definition of the path integral would involve first
defining it on a lattice and then letting the lattice approach the
continuum~\cite{FM87}. Alternatively we can define the path integral by
its perturbative expansion around a saddle point. In both cases, a
finite ultraviolet cut-off $\Lambda$ will be required since the Skyrme
model is non-renormalizable. In the case of chiral perturbation theory
the limit $\Lambda\rightarrow\infty$ can be taken by absorbing the
infinities in the higher order derivative terms, which are neglected at
large distances.

\begin{figure}
\epsfysize=10.0truecm
\epsffile{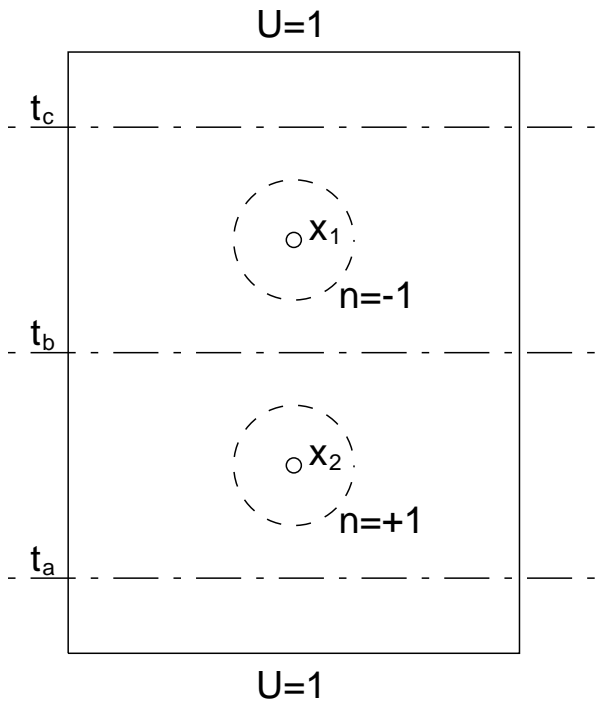}
\caption{To define the propagator, configurations over $R^4\backslash
\{x_1,x_2\}$ are considered. See the discussion in the text.}
\label{fig:sketch}
\end{figure}

In order to clarify the meaning of eq. (\ref{eq:prop}) let us consider 
the soliton propagator
\begin{equation}
G^{-1,+1}(x_1,x_2) = \int {\cal D}[U]
\exp(-\int_{M_{12}} {\cal L}[U]) .
\end{equation}
The path integral is performed over fields  
$U$ which have winding number $+1$ around $x_2$ and $-1$ around $x_1$.
The situation is depicted in figure~\ref{fig:sketch}.

Consider a spatial section at constant time $t=t_a$. By closing this
section at $t=-\infty$ we see that it has topological number 0, that
is, there are no topological excitations at $t=t_a$. The same argument
for a section at $t=t_b$ shows that it has winding number $+1$ and, at
$t=t_c$, the winding number is again $0$. Then, we see that a $B=1$
soliton is created at $x_2$ and destroyed at $x_1$. Furthermore, if we
look at the  time inverted process, we find that section $t=t_b$ has
topological number $-1$,  because definition (\ref{eq:B}) picks up a
minus sign when the normal is reversed. Thus, the figure also describes
an antisoliton propagating backwards in time, that is, at $x_2$ a
soliton is created or equivalently an antisoliton destroyed, the effect
expected from a quantum field operator associated with the antisoliton.
Remember that a field $\psi$ destroys the particle and creates the
antiparticle whereas $\psi^\dagger$ does the opposite.

The following step is to find a saddle point configuration for our path
integral. This can be done proposing a (time-dependent) Skyrme ansatz
\begin{equation}
U=\exp(\mathrm{i} f(u,v) \breve{x}_a \tau^a) ,
\label{eq:ansatz}
\end{equation}
where  $\breve{x}_a=x_a/r$, $r=\sqrt{x_1^2+x_2^2+x_3^2}$,
and $u=e f_{\pi}r$, $v = e f_{\pi}t$ are adimensional.
Minimizing the action, gives an equation for $f(u,v)$: 
\begin{eqnarray}
\left( 2 \sin^2f + u^2\right) \left(\frac{d^2f}{dv^2} +
\frac{d^2f}{du^2} \right) + \sin(2f) \left( \left( 
\frac{df}{dv}\right)^2 + \left(\frac{df}{du}\right)^2 \right) 
+ \nonumber &&\\
+2 u \frac{df}{du} 
- \sin(2f) \left( 1 + \frac{\sin^2f}{u^2}\right) = 0. &&
\label{eq:f}
\end{eqnarray}
The boundary condition $U=1$ at space-time infinity requires that
$f(u,v)\rightarrow 0 $ when $v\rightarrow\pm\infty$ or 
$u\rightarrow\infty$. At $u=0$, for $U$
to be single valued, $f$ must be $0$ or $\pi$. Continuity of $f$ 
(on the manifold $M_{12}$) and the requirement that
$U$ must have topological number $1$ around $x_2$ and $-1$ around $x_1$
leads to
\begin{eqnarray}
f(u,v) &=& 0 \ \ \ \mbox{if} \ \ v\rightarrow \pm\infty\ \ \mbox{or} 
                        \ \ u\rightarrow \infty, \nonumber\\
f(0,v) &=& 0 \ \ \ \mbox{if} \ \ v<ef_\pi x_2^4\ \  
                   \mbox{or}\ \  v>ef_\pi x_1^4,
\label{eq:bcond} \\
f(0,v) &=& \pi \ \ \ \mbox{if} \ \ ef_\pi x_2^4<v<ef_\pi x_1^4. \nonumber
\end{eqnarray}
These boundary conditions respect 
the cylindrical symmetry of the ansatz. In the next section
we will discuss a more general situation.

The numerical solution to eqs. (\ref{eq:f}) and (\ref{eq:bcond}) is
depicted in figure~\ref{fig:numsol} where we see that $f$ (and so the
fields $\phi_a$) is approximately zero before $x_2^4$ or after $x_1^4$
and is almost equal to the static soliton profile in between. This
supports the picture that  a skyrmion is created at $x_2$ and
propagates towards $x_1$ where it is destroyed. Furthermore, since the
solution is almost time independent in the time interval $(x_2^4,x_1^4)$ and
equal to the static profile, the action is approximately given 
by $S=M_{\mathrm{cl}}|x_2-x_1|$, where $M_{\mathrm{cl}}$ is the energy of
the static solution. Hence, the saddle point approximation to the
propagator contains a factor $\exp(-M_{\mathrm{cl}}|x_2-x_1|)$ which is
the exponential behavior of the Euclidean propagator for a particle
with mass $M_{\mathrm{cl}}$. In section 5 we will improve this
approximation, obtaining the correct, spin equal to isospin, particle
spectrum.

\begin{figure}
%\epsfxsize=12.5truecm
%\epsffile{fig2.ps}
\framebox[12truecm]{\rule{0cm}{6truecm} \parbox{10cm}{This figure is available
at http://www.tandar.cnea.edu.ar/preprints/FNT/96/003.tex}}
\caption{Numerical solution for $f(u,v)$ that minimizes the action 
($u=ef_\pi r$, $v=ef_\pi t$). In the figure, $|x_2-x_1| = 6 (ef_\pi)^{-1}$.}
\label{fig:numsol}
\end{figure}

\begin{figure}
\epsfxsize=10truecm
\epsffile{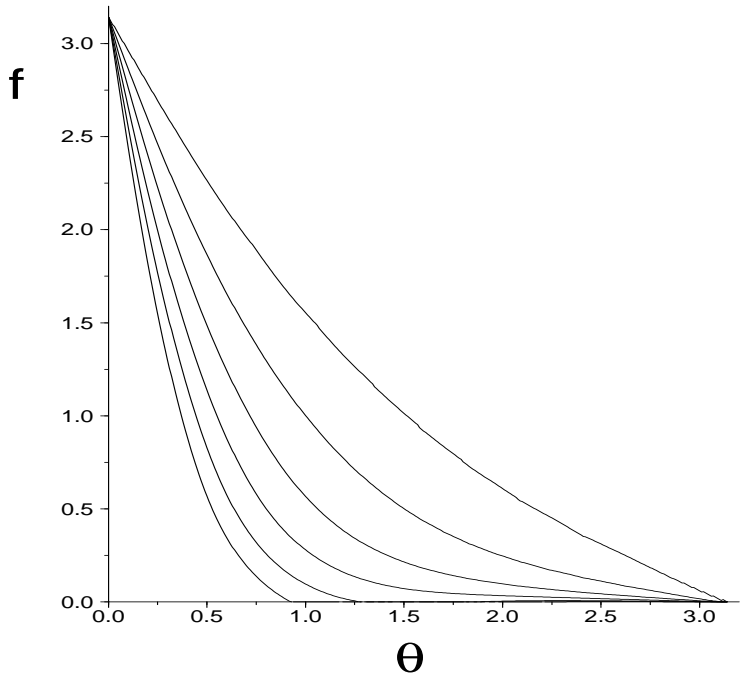}
\caption{$f$ as a function of $\theta$ for different radius, corresponding
(from the upper to the lower curve) to 
$R = 0.5,1.0,1.5,2.0,2.5,3.0$ 
in units of $(ef_\pi)^{-1}$, in the case $|x_2-x_1|=6 (ef_\pi)^{-1}$.}
\label{fig:fvstheta}
\end{figure}

Let us now consider spheres $S_R$ of radius $R<|x_2-x_1|$ around $x_2$.
All these spheres have topological
number $1$ but the topological density, which in this case is given by
$\rho(x_\mu) = B^\mu (x-x_2)^\mu /|x-x_2|$, will be differently distributed.
Replacing the ansatz of eq. (\ref{eq:ansatz}) in the last expression
it follows that the density is only a function of $\theta$, the angle
between $x-x_2$ and the $4$-axis, and is given by
\begin{equation}
\rho(\theta) = -\frac{1}{2\pi^2} 
\frac{\sin^2(f) \partial_\theta f}{\sin^2(\theta)}.
\label{density}
\end{equation}
The numerical solution $f(u,v)$ as a function of $\theta$ can be
obtained, by interpolation, from the formula
\begin{equation}
f(\theta) = f(R\sin(\theta),ef_\pi x_2^4 + R\cos(\theta)).
\end{equation}
The result for several values of $R$ (in units of $(ef_\pi)^{-1}$) is
depicted in figure~\ref{fig:fvstheta}.

In this figure, we see that near the singularity $(R\rightarrow 0)$
$f(\theta)$ tends to $\pi-\theta$, which gives a uniform topological density
(cf. eq. (\ref{density})). As one gets further from the singularity the
topological density concentrates around the North pole of the sphere
giving rise to a soliton moving towards $x_1$.  This is again the
graphical picture corresponding to a soliton appearing around $x_2$.
This behavior of the solution (unlocalization for small $R$, and
localization for large $R$) is in agreement with studies of the static
ansatz on a sphere~\cite{MR86} and is in correspondence with the fact
that at short distances the chiral symmetry is unbroken whereas it is
broken at large distances.

We end this section noticing that the behavior $f(\theta)\approx\pi-\theta$
near $x_2$ is equivalent to 
\begin{equation}
U(x) \approx -X_2^\dagger(x),\ \ \ X_2(x) = \frac{x^4-x_2^4}{|x-x_2|} + 
\mathrm{i}\frac{x^a-x_2^a}{|x-x_2|}\tau_a,
\end{equation}
where $\tau_a$ are the Pauli matrices. A similar analysis around $x_1$,
reveals that $U$ behaves as $X_1$, where $X_1$ is defined as $X_2$, but
with $x_2$ replaced by $x_1$. These behaviors will be useful in the
next section, when discussing the boundary conditions to be imposed on
the fields.

\section{The boundary conditions} \label{sec:bc}

The numerical saddle point obtained in the previous section 
is not invariant under rotations or isorotations, that is, there
are several other saddle points or equivalently, several classical
trajectories.
This is due to the fact that we have not yet specified the
boundary conditions satisfied by the field around $x_1$ and $x_2$.

In order to do so, we extract two small balls $B^4_1$ and $B^4_2$ of radius 
$\epsilon\rightarrow 0$ 
around $x_1$ and $x_2$ respectively. Then, we need to specify the values 
taken by the field $U$ on the surfaces $S_1^3=\partial B^4_1$ and 
$S^3_2=\partial B^4_2$ of these balls, satisfying
the condition that $U$ must have topological number $+1$ around $x_2$ 
and $-1$ around $x_1$. 

In principle, any function $U:S^3 \rightarrow \mathrm{SU(2)}$ can be
chosen, but we will restrict ourselves to a minimum set of boundary
conditions which close under Lorentz and isospin transformations. At
the end of section \ref{sec:Gf} we saw that near $x_2$ the numerical
solution behaves as $-X_2^\dagger$, resulting in a uniform topological
density. Performing Lorentz transformations we obtain, around the
singularity, the behaviors $S_1(-X_2)^\dagger S_2^\dagger$
($S_1,S_2\in\mathrm{SU(2)}$, see eq. (\ref{eq:lt}) below) without
altering the uniform topological density. This suggest that all the
boundary conditions $S_1(-X_2)^\dagger S_2^\dagger$ should be treated
on an equal footing to obtain a representation of Lorentz and isospin
symmetries.  Thus, the boundary conditions are taken to be
\begin{eqnarray}
\left. U \right|_{S^3_1} &=& A_1 X_1 B_1^\dagger, \nonumber\\
\left. U \right|_{S^3_2} &=& A_2 (-X_2)^\dagger B_2^\dagger, 
\label{eq:bc}
\end{eqnarray}
where $A_1,B_1,A_2,B_2$ belong to $\mathrm{SU(2)}$ and $X_1,X_2$ are the 
$\mathrm{SU(2)}$ matrices defined by
\begin{equation}
X_{1,2}(x) = \frac{x^4-x_{1,2}^4}{|x-x_{1,2}|} + 
\mathrm{i}\frac{x^a-x_{1,2}^a}{|x-x_{1,2}|}\tau_a,
\end{equation}
As an aside remark, note that, in terms of the vector field 
$L_\mu = U^\dagger\partial_\mu U$, these conditions look like the 
insertion of a zero-radius instanton (anti-instanton) at 
$x_1$ ($x_2$).\footnote{ In our case we do not have
a gauge theory, but one can define the Green functions in terms of a
field $U$ coupled to a singular $L_{\mu}$ as described in~\cite{FM90}}

Each pair $A_i,B_i$ determines a map $S^3_i\rightarrow \mathrm{SU(2)}$. 
In fact, as exactly the same map is determined by $-A_i,-B_i$, the boundary
condition is defined by a point in the manifold $\mathrm{SU(2)}\times
\mathrm{SU(2)}/Z_2$. These boundary conditions represent the internal 
degrees of freedom of the skyrmion fields.
  
Given two points $p_1$ and $p_2$ on $\mathrm{SU(2)}\times
\mathrm{SU(2)}/Z_2$ we define a propagator
\begin{equation}
G(x_1,p_1;x_2,p_2) =\lim_{\epsilon\rightarrow 0}
 \int {\cal D}[U] \exp\left(-\int_{M_{12}} {\cal L}[U]+ 2 C_\epsilon\right),
\label{eq:propG}
\end{equation}
where $U$ at $S^3_1$  is given by $p_1$ and at $S^3_2$ by $p_2$, 
according to eq. (\ref{eq:bc}) (at space-time infinity we take $U=1$, 
the vacuum configuration).
$C_\epsilon = -(3\pi^2/e^2) \log(\epsilon)$ is added to cancel
the infinite contribution to the action due to the singularity and
can be thought of as wave function renormalization of the skyrmion fields.
Including higher order derivative terms in the action would produce worse
singularities and $C_\epsilon$ should be properly redefined;
the inclusion of, say, the sixth order term $-\epsilon_6 B^\mu B_\mu$ 
requires that $C_\epsilon = -(6\pi^2/e^2)\log(\epsilon)- 
\epsilon_6 (2\pi^2\epsilon^2)^{-1}$. Of course, at short distances,
the most significant terms are precisely those 
neglected at large distances, reflecting the fact that near the 
singularities the lagrangian
in eq. (\ref{eq:lag}) is not adequate. However this does not modify the
symmetry arguments and, far from the singularities, the classical 
solution is insensitive to the higher order terms.

Now, we will show that the imposed boundary conditions lead to a
natural discussion of the different possible quantizations of the
Skyrme model.
Up to now, as $M_{12} = (R^4\cup\{\infty\})\backslash
 \{x_1,x_2\}$ is
contractible to $S^3$, the configurations $U:M_{12}\rightarrow \mathrm{SU(2)}$ 
are classified by an integer. However, when we fix the initial
and final points we have a map
\begin{equation}
U: S^3\times[0,1] \rightarrow \mathrm{SU(2)},
\end{equation}
because $M_{12}$ can be deformed to a cylinder with base $S^3$. Here,
$\partial B^4_1$ and $\partial B^4_2$ are the bases of the cylinder and
the parameter in $[0,1]$ represents the continuous deformation of
$\partial B^4_2$ into $\partial B^4_1$, passing through sections
between $x_1$ and $x_2$ (remember that space-time infinity is identified).
Two such maps are not necessarily homotopic. The situation is similar
to that of a plane with a hole. Paths with free end points are all
homotopic but paths with fixed end points are classified by
$\Pi_1(S^1)=Z$.

In our case suppose we give two configurations
\begin{equation}
U_1,U_2 : S^3\times[0,1] \rightarrow \mathrm{SU(2)},
\end{equation}
which satisfy the same boundary conditions at $0$ and $1$.
They can be used to define a map $U:S^3\times S^1 \rightarrow \mathrm{SU(2)}$
as:
\begin{equation}
U(X,t)=\left\{\begin{array}{lll}U_1(X,2t)&\ \mbox{if}& \ t\in [0,1/2]\\
                              U_2(X,2(1-t))&\ \mbox{if}& \ t\in[1/2,1].
              \end{array} \right.
\label{eq:U12}
\end{equation}
It is easy to verify that $U$ is continuous at $1/2$ and
periodic ($U(X,0)=U(X,1)$), provided that $U_1$ and $U_2$ satisfy 
the same boundary conditions at $t=0,1$. This map is the analog
of the closed path which can be constructed from two paths with the 
same end points, in the simple example of the plane with a hole.

Two configurations $U_1$ and $U_2$ are homotopic if the map 
$U:S^3\times S^1 \rightarrow \mathrm{SU(2)}$ can be
extended to a map $S^3\times D_2 \rightarrow \mathrm{SU(2)}$, where
$D_2$ is a two dimensional disc $(\partial D_2=S^1)$. In general, 
there is an obstruction to
such an extension given by $\Pi_4(\mathrm{SU(2)})=Z_2$~\cite{Wi83},
this means that there are two different homotopy classes. A way of
distinguishing them is to trivially embed $\mathrm{SU(2)}$ into
$\mathrm{SU(3)}$ and evaluate the Wess-Zumino-Witten term. This gives
$0$ for the trivial class and $\pi$ for the non-trivial one. If we were
considering $\mathrm{SU(3)}$ solitons all the paths would be homotopic
and the  Wess-Zumino-Witten term would give a different contribution
for each path \cite{Wi83}. We would like to point out that the need of
considering an extension of a map $U:S^3\times S^1 \rightarrow
\mathrm{SU(2)}$ to a map $U:S^3\times D_2 \rightarrow \mathrm{SU(2)}$
to study the different possible inequivalent quantizations of the
Skyrme model, which in~\cite{Wi83} was proposed by imposing periodic
boundary conditions to calculate the partition function, here
also arises naturally from the Fr\"ohlich and Marchetti's formalism
supplemented with fixed boundary conditions around the singularities.

The path integral defining the propagator can then be split into two
integrals, one over each homotopy class. A phase equal to $0$ or $\pi$
can be assigned to the non-trivial class, which corresponds to adding
or subtracting the two contributions:
\begin{equation}
G(x_1,p_1;x_2,p_2) = G^{(1)}(x_1,p_1;x_2,p_2)\pm G^{(2)}(x_1,p_1;x_2,p_2). 
\end{equation}
The plus (minus) sign will correspond to quantizing the soliton as a 
boson (fermion) (see section \ref{sec:Sp}). 

Now, suppose that we have a path in some topological class which 
ends at a boundary condition associated to $A_2,B_2$; 
if we adiabatically move this last boundary condition, leaving  
$A_1,B_1$ fixed, to get a boundary condition associated to 
$-A_2,-B_2$ (which represents the same point 
as $A_2,B_2$ and  $-A_2,-B_2$ are identified),
the new path will be in a different topological class. 
This is due to the fact that 
the new path can be obtained by composing the original path with a 
``closed'' path, based on $\partial B^4_2$, that starts at 
$(A_2,B_2)$ and ends at $(-A_2,-B_2)$. This last path is non-trivial 
and so the original path will change from homotopy class.  
Thus, changing adiabatically the final condition from $(A_2,B_2)$
to $(-A_2,-B_2)$ interchanges $G^{(1)}$ and $G^{(2)}$.

This result allows us to continuously extend the boundary 
conditions from $\mathrm{SU(2)}\times \mathrm{SU(2)}/Z_2$ to 
$\mathrm{SU(2)}\times \mathrm{SU(2)}$, defining a new function 
$\tilde{G}$ according to
\begin{eqnarray}
\tilde{G}(x_1, A_1, B_1;x_2, A_2, B_2) &=&
\tilde{G}(x_1,-A_1,-B_1;x_2,-A_2,-B_2) \nonumber\\
&=&  G^{(1)}(x_1,p_1;x_2,p_2), \nonumber\\
\tilde{G}(x_1, A_1, B_1;x_2,-A_2,-B_2) &=& 
\tilde{G}(x_1,-A_1,-B_1;x_2, A_2, B_2) \nonumber\\
&=&  G^{(2)}(x_1,p_1;x_2,p_2), 
\label{eq:Gtilde}
\end{eqnarray} 
where $A_{1,2},B_{1,2}$ and $-A_{1,2},-B_{1,2}$ both correspond to
$p_{1,2}$ in $\mathrm{SU(2)}\times \mathrm{SU(2)}/Z_2$ and we take
the definitions to be valid when $A_1,A_2 > 0$ \footnote{We define
$A>0$, $A\in \mathrm{SU(2)}$ when $A=a_0 + \mathrm{i}
\vec{a}.\vec{\tau}, (a_0^2+\vec{a}^2=1)$ is written with $a_0>0$.}.  In
terms of $\tilde{G}$ the original propagator can be written as
\begin{equation}
G(x_1,p_1;x_2,p_2) = 
\tilde{G}(x_1, A_1, B_1;x_2, A_2, B_2)\pm
\tilde{G}(x_1,A_1,B_1;x_2,-A_2,-B_2).
\label{eq:GGtilde}
\end{equation}
In the subsequent sections we will analyze the properties of $\tilde{G}$.

To close this section let us point out that
the propagator $G$ can be associated with a field propagator
\begin{equation}
G(x_1,p_1;x_2,p_2) = \langle 0|T\{\psi^1_{A_1,B_1}(x_1)
\psi^{-1}_{A_2,B_2}(x_2)\}|0\rangle.
\label{Greenfield}
\end{equation}
Similarly, Green functions for any number ($p$)
of operators $\psi^n_{A,B}(x)$, and operators $\phi_a(y)$, can be defined 
(with the restriction $n_i=\pm 1$) according to:
\begin{eqnarray}
\lefteqn{\langle 0| T\{\psi_{A_1,B_1}^{n_1}(x_1)\ldots 
\psi_{A_p,B_p}^{n_p}(x_p) \phi_{a_1}(y_1)\ldots \phi_{a_r}(y_r)
\} |0\rangle =}\nonumber \\ 
&&\lim_{\epsilon\rightarrow 0} \int
{\cal D}[U] \phi_{a_1}(y_1)\ldots \phi_{a_r}(y_r)
\mathrm{e}^{-S[U] + p C_\epsilon},
\label{mixed}
\end{eqnarray}
together with the boundary conditions,
\begin{eqnarray}
\left. U \right|_{S_i} &=& \left\{ 
\begin{array}{lll}
A_i X_i B_i^\dagger & \mbox{if}& n_i=-1 \\
A_i (-X_i)^\dagger B_i^\dagger & \mbox{if}& n_i=+1 
\end{array} \right. \\ 
X_i &=& 
\frac{x^4-x^4_i}{|x-x_i|} + 
\mathrm{i} \frac{x^a-x^a_i}{|x-x_i|} \tau_a \in \mathrm{SU(2)},\nonumber
\end{eqnarray}
for small spheres $S_i$ around $x_i$ 
($C_\epsilon = -(3 \pi^2/e^2)\log(\epsilon)$).

If $n>1$, the operator $\psi^n(x)$ could be defined as a
suitable limit of $\psi^1(y_1)\ldots \psi^1(y_n)$ when
$y_i\rightarrow x$. However we will not pursue this point 
here but instead, from now on we will consider only $\psi^{\pm 1}$ fields 
which will be denoted as $\psi^{+1} = \psi$, $\psi^{-1}=\bar{\psi}$. 

\section{Symmetries of the propagator} \label{sec:sotp}

The symmetries of the propagator can be found by performing different
changes of variables, within the path integral, which are invariances of the
action. In this way we learn how this symmetries act on the internal variables
$A_i$, $B_i$.
\begin{description}
\item[Translations]
The change of variable $U(x) \rightarrow U(x+a)$ does not affect the
action but affects the boundary conditions. Trivially, it follows that
\begin{equation}
\tilde{G}(x_1 + a,A_1,B_1;x_2 + a ,A_2,B_2) = 
\tilde{G}(x_1    ,A_1,B_1;x_2     ,A_2,B_2).
\end{equation}
\item[Isorotations]
With the change of variables $\tilde{U} = A U A^\dagger$ , 
$(A\in \mathrm{SU(2)}$) we obtain
\begin{eqnarray}
S[AUA^\dagger] &= & S[U], \nonumber\\
A 1 A^\dagger &=&1 \ \ \ \mbox{at}\  \infty, \nonumber\\
\left. \tilde{U} \right|_{S^3_1} &=& 
A \left. U \right|_{S^3_1}A^\dagger = AA_1 X_1 B_1^\dagger A^\dagger =
(AA_1) X_1 (AB_1)^\dagger, \nonumber\\
\left. \tilde{U} \right|_{S^3_2} &=& 
A \left. U \right|_{S^3_2}A^\dagger = AA_2 (-X_2)^\dagger B_2^\dagger A^\dagger =
(AA_2) (-X_2)^\dagger (AB_2)^\dagger. 
\end{eqnarray}
Thus
\begin{equation}
\tilde{G}(x_1,AA_1,AB_1;x_2,AA_2,AB_2) = 
\tilde{G}(x_1, A_1, B_1;x_2, A_2, B_2).
\label{eq:isorot}
\end{equation}
\item[Lorentz transformations]
To describe particles of half-integer spin we must consider  
the universal covering of the Euclidean Lorentz group, namely
$\mathrm{SU(2)}\times \mathrm{SU(2)}$. To each
pair of matrices $S_1,S_2\in \mathrm{SU(2)}$  corresponds a
matrix $\Lambda \in \mathrm{SO(4)}$.
 The transformation $x\rightarrow x'=\Lambda x$ can
be written in matrix form
\begin{equation}
X \rightarrow  X'=S_1XS_2^\dagger , \ \ \ 
 X=\frac{x^4}{|x|}+\mathrm{i}\frac{x^a}{|x|}\tau_a \in \mathrm{SU(2)}.
\label{eq:lt}
\end{equation}
Now let us perform the change of variables 
\begin{equation}
U(x)\rightarrow \tilde{U}(x)=U(\Lambda^{-1}x),
\end{equation}
which leaves the action invariant.
The boundary conditions transform as
\begin{eqnarray}
\left. \tilde{U}(X') \right|_{S'^3_1} &=& 
\left. U(X) \right|_{S^3_1} = A_1X_1B_1^\dagger =  
A_1S_1^\dagger X'_1 S_2B_1^\dagger =
(A_1S_1^\dagger) X_1 (B_1S_2^\dagger)^\dagger, \nonumber\\
\left. \tilde{U}(X') \right|_{S'^3_2} &=& 
\left. U(X) \right|_{S^3_2} = A_2(-X_2^\dagger)B_2^\dagger =  
(A_2S_2^\dagger) (-X_2^{'\dagger}) (B_2S_1^\dagger)^\dagger,
\end{eqnarray}
which gives
\begin{equation}
\tilde{G}(\Lambda x_1, A_1S_1^\dagger, B_1S_2^\dagger ; 
  \Lambda x_2, A_2S_2^\dagger, B_2S_1^\dagger) =
\tilde{G}(x_1,A_1,B_1;x_2,A_2,B_2).
\label{eq:lorentz}
\end{equation}
\item[Parity]
The pion is pseudo-scalar, so under parity $U$ transforms
as
\begin{equation}
U(x^4,\vec{x})\rightarrow \tilde{U}(x^4,\vec{x}) = 
U^\dagger(x^4,-\vec{x}).
\end{equation}
As a spatial reflection maps 
$X_{1,2}\rightarrow X_{1,2}^\dagger$, a parity transformation
simply interchanges $A$ and $B$ in the boundary conditions.
So, parity invariance of the action implies that
\begin{equation}
\tilde{G}(x_1^4,\vec{x}_1, A_1, B_1; x_2^4,\vec{x}_2, A_2, B_2) =
\tilde{G}(x_1^4,-\vec{x}_1, B_1, A_1; x_2^4,-\vec{x}_2, B_2, A_2).
\label{eq:parity}
\end{equation}

\end{description}

Let us proceed now to find how these symmetries constraint the form 
of the propagator.
Because of translational invariance $\tilde{G}$ depends on the difference
$x_1-x_2$.
The vector $x_1-x_2$ can be split into a radial and an angular part
as follows
\begin{equation}
x_1-x_2 \rightarrow \left\{ \begin{array}{l}|x_1-x_2|\\
         X_{12} = \frac{(x_1^4-x_2^4)}{|x_1-x_2|} +\mathrm{i} 
                  \frac{(x_1^a-x_2^a)}{|x_1-x_2|} \tau_a
         \in \mathrm{SU(2)}.
         \end{array} \right.
\end{equation}

$\tilde{G}$ is a smooth function of $X_{12},A_1,B_1,A_2,B_2$.
Any such function can be decomposed as a linear combination of
the matrix elements $D^j_{\sigma\sigma'}$. This can
be seen from group theory or because
the $D^j$ matrices are the spherical harmonics on $S^3$.
Thus, we expand $\tilde{G}$ as
\begin{eqnarray}
\lefteqn{\tilde{G}(|x_1-x_2|,X_{12},A_1,B_1,A_2,B_2)=
\sum_{j_i,\  |\sigma_i|,|\sigma'_i|\le j_i}
\eta(j_i,\sigma_i,\sigma'_i;|x_2-x_1|)} \nonumber \\ 
&&
D^{j_1}_{\sigma_1\sigma'_1}(A_1)
D^{j_2}_{\sigma_2\sigma'_2}(B_1)
D^{j_3}_{\sigma_3\sigma'_3}(A_2)
D^{j_4}_{\sigma_4\sigma'_4}(B_2)
D^{j_5}_{\sigma_5\sigma'_5}(X_{12}).
\end{eqnarray}
where $j$ runs over $\half Z_{\ge 0}$, i.e. integer and half-integer 
representations. 
From definition (\ref{eq:Gtilde}) we see that
\begin{eqnarray}
&&\tilde{G}(x_1, A_1, B_1;x_2, A_2, B_2) =
\tilde{G}(x_1,-A_1,-B_1;x_2,-A_2,-B_2) \nonumber\\
&&\Rightarrow
(-)^{2(j_1+j_2+j_3+j_4)}=1, 
\label{eq:sign}
\end{eqnarray}
where the implication follows from the formula $D^j(-A) = (-)^{2j} D^j(A)$.

Performing independent variations of $A$, $S_1$, $S_2$ in eqs.
(\ref{eq:isorot}), (\ref{eq:lorentz}) and using standard arguments
about the coupling of $\mathrm{SU(2)}$ representations one finds that the $\eta$ 
indices $(\sigma_1\sigma_2\sigma_3\sigma_4)$ 
are coupled to zero (isorotations), as well as the indices
$(\sigma'_2,\sigma'_3,\sigma'_5)$ and $(\sigma'_1,\sigma'_4,-\sigma_5)$
(Lorentz transformations). Thus,
the most general $\tilde{G}$ satisfying all these symmetries is 
\begin{eqnarray}
\lefteqn{\tilde{G}(|x_1-x_2|,X_{12},A_1,B_1,A_2,B_2)=} \nonumber\\ 
&&\sum_{\mbox{all}\ \sigma,\sigma',j,m,I} (2I+1)^2 
C_{\sigma'_1\bar{\sigma}_1} C_{\bar{\sigma}_4\sigma'_4} C_{m_2m_1}
\eta_{j_1j_2j_3j_4j_5I}(|x_2-x_1|) \nonumber\\
&&\left(\begin{array}{lll}j_1&j_4&j_5\\
\bar{\sigma}_1&\bar{\sigma_4}&\sigma_5\end{array}\right)
\left(\begin{array}{lll}j_3&j_2&j_5\\
\sigma'_3&\sigma'_2&\sigma'_5\end{array}\right)
\left(\begin{array}{lll}j_1&j_2&I\\
\sigma_1&\sigma_2&m_1\end{array}\right)
\left(\begin{array}{lll}j_3&j_4&I\\
\sigma_3&\sigma_4&m_2\end{array}\right) \nonumber\\
&& D^{j_1}_{\sigma_1\sigma'_1}(A_1)
D^{j_2}_{\sigma_2\sigma'_2}(B_1)
D^{j_3}_{\sigma_3\sigma'_3}(A_2)
D^{j_4}_{\sigma_4\sigma'_4}(B_2)
D^{j_5}_{\sigma_5\sigma'_5}(X_{12}),
\label{eq:prop_eta}
\end{eqnarray}
where use was made of the Wigner's $3j$ symbols (which couple to 
zero three angular momenta) and of the matrices 
$C_{mm'} = (-)^{j+m} \delta_{m,-m'}$. Useful properties of $C$ are
$C=C^*,  C^{-1} = C^t = (-)^{2j} C$.

Parity conservation implies $\eta_{j_1j_2j_3j_4j_5I} =
\eta_{j_2j_1j_4j_3j_5I}$ (cf. eq. (\ref{eq:parity})).

The transformation laws of this section can also be defined over
the fields $\psi_{A,B}$,$\bar{\psi}_{A,B}$ by means of unitary
operators $\mathrm{U}$ satisfying:

Isorotations:
\begin{eqnarray}
\mathrm{U}_A \psi_{A_1,B_1}(x) \mathrm{U}^\dagger_A &=& 
 \psi_{AA_1,AB_1}(x), \nonumber \\
\mathrm{U}_A \bar{\psi}_{A_1,B_1}(x) \mathrm{U}^\dagger_A &=& 
 \bar{\psi}_{AA_1,AB_1}(x).
\label{eq:IAB}
\end{eqnarray}
Lorentz transformations:
\begin{eqnarray}
\mathrm{U}_{\Lambda}\psi_{A_1,B_1}(x)\mathrm{U}^\dagger_\Lambda =
 \psi_{A_1S_1^\dagger,B_1S_2^\dagger}(\Lambda x), \nonumber \\ 
\mathrm{U}_{\Lambda}\bar{\psi}_{A_2,B_2}(x)\mathrm{U}^\dagger_\Lambda =
 \psi_{A_1S_2^\dagger,B_1S_1^\dagger}(\Lambda x),
\label{eq:LAB}
\end{eqnarray}
where $\Lambda \equiv (S_1,S_2)$.

These transformations suggest changing to a basis of fields which 
transform under a finite dimensional representation of the Lorentz
and isospin groups. Finite dimensional irreducible representations of the 
$\mathrm{SU(2)}\times \mathrm{SU(2)}$ 
group are labeled by two numbers $j_1,j_2 \in \half Z_{\ge 0}$. If 
$\Lambda \equiv (S_1,S_2)$ the representation $(j,0)$ is given by
${\cal D}^{j}(\Lambda) = D^{j}(S_1)$ and $(0,j)$ by 
$\bar{{\cal D}}^{j}(\Lambda) = D^{j}(S_2)$.  The matrices $D^{j}$ are 
the usual rotation matrices representing $\mathrm{SU(2)}$ and so, unlike the 
Minkowski case, this representations are unitary but not-faithful. 
Finally the $(j_1,j_2)$ representations are given by
$(j_1,0)\otimes (0,j_2)$.  
The fields transforming under these representations are given by:
\begin{eqnarray}
\psi^{j_1j_2I}_{\sigma_1\sigma_2m_I}(x) &=& 
Z^{-1}_{j_1j_2I} 
C_{\sigma'_1\sigma_1} C_{\sigma'_2\sigma_2} 
\left(\begin{array}{lll}j_1&j_2&I\\
\sigma''_1&\sigma''_2&m_I\end{array}\right) \nonumber\\
&&
\int_{A,B} D^{j_1}_{\sigma''_1\sigma'_1}(A)
D^{j_2}_{\sigma''_2\sigma'_2}(B)
\psi_{A,B}(x),
\label{eq:psi}
\end{eqnarray}
and the same formula for $\bar{\psi}$.
The integration over $\mathrm{SU(2)}$ is performed using the group invariant 
measure and $Z_{j_1j_2I}$ is a normalization factor to be chosen later.
The transformations (\ref{eq:IAB}), (\ref{eq:LAB}) now read:
\begin{eqnarray}
\mathrm{U}_A \psi^{j_1j_2I}_{\sigma_1\sigma_2m_I}(x)
\mathrm{U}^\dagger_A &=& D^I_{m'_Im_I}(A) 
\psi^{j_1j_2I}_{\sigma_1\sigma_2m'_I}(x),\nonumber\\ 
\mathrm{U}_A \bar{\psi}^{j_1j_2I}_{\sigma_1\sigma_2m_I}(x)
\mathrm{U}^\dagger_A &=& D^I_{m'_Im_I}(A) 
\psi^{j_1j_2I}_{\sigma_1\sigma_2m'_I}(x), \nonumber\\
\mathrm{U}_{\Lambda} \psi^{j_1j_2I}_{\sigma_1\sigma_2m_I}(x)
\mathrm{U}^\dagger_{\Lambda} &=&
{\cal D}^{j_1}_{\sigma_1\sigma'_1}(\Lambda^{-1})
\bar{{\cal D}}^{j_2}_{\sigma_2\sigma'_2}(\Lambda^{-1})
\psi^{j_1j_2I}_{\sigma'_1\sigma'_2m_I}(\Lambda x), \nonumber\\
\mathrm{U}_{\Lambda} \bar{\psi}^{j_1j_2I}_{\sigma_1\sigma_2m_I}(x)
\mathrm{U}^\dagger_{\Lambda} &=&
\bar{{\cal D}}^{j_1}_{\sigma_1\sigma'_1}(\Lambda^{-1})
{\cal D}^{j_2}_{\sigma_2\sigma'_2}(\Lambda^{-1})
\bar{\psi}^{j_1j_2I}_{\sigma'_1\sigma'_2m_I}(\Lambda x),
\label{eq:LABpsi}
\end{eqnarray}
which means that $\psi^{j_1j_2I}_{\sigma_1\sigma_2m_I}(x)$
transforms under the Lorentz group representation $(j_1,j_2)$
and $\bar{\psi}^{j_1j_2I}_{\sigma_1\sigma_2m_I}(x)$ under 
the representation $(j_2,j_1)$. In addition, both fields have isospin $I$.

Using eq. (\ref{eq:prop_eta}) and definition (\ref{eq:psi}) we can obtain 
the general form of the propagator 
\begin{eqnarray}
\langle0|T\{\bar{\psi}^{j_3j_4I'}_{\sigma_3\sigma_4m'_I}
\psi^{j_1j_2I}_{\sigma_1\sigma_2m_I}\}|0\rangle &=&
Z_{j_1j_2I}^{-1} Z_{j_3j_4I'}^{-1} \delta_{II'} C_{m'_Im_I}
D^{j_1}_{\sigma_1\sigma'_1}(X_{12})
D^{j_4}_{\sigma_4\sigma'_4}(X_{12}) \nonumber \\
&&
\sum_{j_6}\left(\begin{array}{lll}j_1&j_2&j_6\\
\sigma'_1&\sigma_2&\sigma_6\end{array}\right)
\left(\begin{array}{lll}j_3&j_4&j_6\\
\sigma_3&\sigma'_4&\sigma'_6\end{array}\right) 
C_{\sigma'_6\sigma_6} \bar{\eta}_{j_1j_2j_3j_4j_6I} \nonumber \\
\label{eq:prop_etab}
\end{eqnarray}
where $\bar{\eta}$ can be written in terms of $\eta$ and the $6j$ symbols~\cite{E60} as 
\begin{equation}
\bar{\eta}_{j_1j_2j_3j_4j_6I} = \sum_{j_5} 
\eta_{j_1j_2j_3j_4j_5I} (-)^{2j_5} 
\left\{\begin{array}{lll}j_1&j_2&j_6\\j_3&j_4&j_5\end{array}\right\}.
\end{equation}

The spins associated with these fields are obtained by composing $j_1$
and $j_2$ and so, in principle, they range from $|j_2-j_1|$ to
$j_1+j_2$. The propagated spins are determined by the propagator in
(\ref{eq:propG}). In the next section we will show that
(\ref{eq:propG}) has a new symmetry which leads to the propagation of
particles with spin equal to the isospin $I$.

\section{Skyrmion propagator} \label{sec:Sp}

Now, we will approximately evaluate the propagator 
in the large $|x_2-x_1|$ limit. Performing a Lorentz
transformation, the $4$-axis can always be taken along $x_2-x_1$,
i.e. $X_{12}=1$. 
By means of eq. (\ref{eq:prop_etab}) we can extend this result to 
arbitrary values of $X_{12}$.

Having fixed $X_{12}$, we will analize what is expected for
different values of $A_{1,2}$ and $B_{1,2}$.

If $A_1 = B_1 = A_2 = B_2 = 1$, the situation is that of
section~\ref{sec:Gf} where we evaluated (numerically) the saddle point.
This saddle point will be the base for our analysis and will
be denoted as $U_0(t,\vec{x})$. One should remember that it behaves
as $-X_2^\dagger$ around $x_2$ and as $X_1$ around $x_1$. Note also that 
$U_0(t,\vec{x})$ (as a function of
$\vec{x}$), in the time interval between $x^4_2$ and $x^4_1$, is almost 
equal to the static soliton profile, which will be denoted as 
$U_{\mathrm{SK}}(\vec{x})$.

Suppose now that $A_2 = B_2 = A$, $A_1 = B_1 = 1$. This corresponds to
performing an isorotation in the vicinity of $x_2$. We know that if $U$
is a solution to the equation of motion, then, $AUA^\dagger$ (with
constant $A$) is also a solution, that is, the function
$AU_0A^{\dagger}$ satisfies the equations of motion around $x_2$. It
also satisfies the boundary condition on $\partial B^4_2$. Moreover,
this behavior can be matched with the behavior around $x_1$ by means of
a time dependent ansatz of the form
\begin{equation}
U(t,\vec{x}) = A(t) U_0 A^\dagger(t),\ \ 
A(x_2^4)=A,\ A(x_1^4)=1, 
\label{srotation}
\end{equation}
which satisfies $U=1$ at infinity; upon substitution in the action we
obtain an equation for $A(t)$. The corresponding solution will
represent, through eq. (\ref{srotation}), the creation at $x_2$ of a
skyrmion isorotated in $A$, with profile $AU_{\mathrm{SK}}A^\dagger$,
which slowly rotates until it is annihilated unrotated at $x_1$. This
solution is expected to give a good approximation to the saddle point.
More generally, for $A_2 = B_2$, $A_1 = B_1$ the saddle point will be
approximately given by minimizing the action with respect to the ansatz
(\ref{srotation}) with $A(x_2^4)=A_2$, $A(x_1^4)=A_1$.  This is by now
a standard calculation in skyrmion physics which gives free motion over
$\mathrm{SU(2)}$ with a moment of inertia $\Im$, obtained from
$U_{\mathrm{SK}}(\vec{x})$~\cite{Wi79}.  A better approximation to the
field propagator can be done by integrating over all possible $A(t)$
giving the free propagator over $\mathrm{SU}(2)$, from $A_2$ to $A_1$,
which corresponds to quantizing the isorotational (or rotational)
zero-modes.

Now, we consider the case $A_2\neq B_2$, $A_1 = B_1 = 1$. In this case
we also have that $A_2U_0 B_2^\dagger$ is a solution in the vicinity of
$x_2$ that satisfies the correct boundary condition at $\partial B^4_2$
but here, a time dependent ansatz does not suffices; 
for such an ansatz, the behavior $A_2U_0 B_2^\dagger$ around the vicinity 
of $x_2$ would be extended to
infinity on the hyperplanes $x^4 = \mathrm{const.}\approx x_2^4$ to a 
value $A_2B_2^\dagger\neq 1$ that does not match the required asymptotic 
behavior of the fields.  Recall that, around $x_2$, $U_0$ is unlocalized 
for $|x-x_2|<r_l$, while it is localized on the North pole for
$|x-x_2|>r_l$ (see the discussion at the end of section~\ref{sec:Gf}).
Then, if we start with a function such that, very close to
$|x-x_2|=\epsilon$, behaves as the solution $A_2U_0 B_2^\dagger$ and
this behavior persists up to the localization radius, the topological
density will not be localized on the North pole of the sphere
$|x-x_2|=r_l$, but on the direction $x-x_2$ satisfying $X_2=
B_2^{\dagger}A_2$. This behavior represents a soliton that starts
moving on a direction other than towards $x_1$. Therefore, the matching
with the behavior around $x_1$ where the soliton is required to be
annihilated will lead to a function with action larger than the action
associated with a function representing a soliton that starts moving
towards $x_1$, that is, we expect that the configuration which minimizes 
the action behaves, in the unlocalization region around $x_2$, as 
\begin{equation}
U(x) = A(\tau) U_0 B(\tau)^\dagger ,\ \ \ A(\epsilon)=A_2,\ B(\epsilon)=B_2,
\ \ \ A(r_l) = B(r_l) = C_2
\end{equation}
where $\tau=|x-x_2|$, $\epsilon$ is the radius of the small sphere we
extracted around $x_2$ and $r_l\sim (ef_\pi)^{-1}$. Note that this
behavior, at $\tau=r_l$, correctly localizes the topological density on
the North pole, though isorotated by $C_2$. The value of $C_2$ must be
chosen so as to minimize the generated action in the unlocalization
region. Loosely speaking, $(C_2,C_2)$ will be the closest rotation to
the Lorentz transformation $(A_2,B_2)$.  A more quantitative answer is
obtained by inserting the above ansatz in the action, leading to
\begin{eqnarray}
S &=& 4\pi^2\int_{\epsilon<\tau<r_l} \mathrm{d}\tau \, h(\tau) \mathrm{Tr}
\left(\dot{A}\dot{A}^\dagger +
\dot{B}\dot{B}^\dagger\right), \nonumber\\  
h(\tau) &=&  \frac{f_\pi^2}{8} \tau^3 + \frac{\tau}{4e^2} 
- \frac{\epsilon_6}{16\pi^2}\frac{1}{\tau} ,
\label{eq:SAB}
\end{eqnarray}
where the dots represent derivation with respect to $\tau$, and we made the
approximation $U_0\simeq -X_2^\dagger$.
After a change of variables $u=u(\tau)$ such that 
$\dot{u} = 1/h(\tau)$, which cancels the factor $h(\tau)$, 
the expression (\ref{eq:SAB}) represents
free motion over $\mathrm{SU(2)}\times\mathrm{SU(2)}$ for $(A(u),B(u))$. For small 
$\tau$, the solution to $\dot{u}=1/h(\tau)$ is 
$u(\tau) \simeq - 8 \pi^2 \tau^2/\epsilon_6$.

At $u\sim u(r_l)$ we must have $A(u)=B(u)=C_2$, then the action is equal to
\begin{eqnarray}
S &=& 8\pi^2 \frac{s_{A_2C_2}^2 + s_{B_2C_2}^2}{u(r_l)-u(\epsilon)},\nonumber\\
s_{AB} &=& 2 \mathrm{Arccos}(\half\mathrm{Tr}(AB^\dagger)).
\end{eqnarray}
Where $s_{AB}$ is the distance between $A$ and $B$ measured
with the standard metric of $\mathrm{SU(2)}$. For $A_2$, $B_2$ fixed, a 
calculation reveals that this action is minimized by 
$C_2=B_2\sqrt{B_2A_2^\dagger}$. Note that in $\mathrm{SU(2)}$, the
square root is well defined up to a sign\footnote{except
for the square root of minus the identity since any matrix 
$A=i\vec{a}.\vec{\tau}, \vec{a}.\vec{a}=1$, satisfies $A^2=-1$.}.
This part of the saddle point action contributes to the
propagator with a factor 
$\exp(-(s_{A_2C_2}^2 + s_{B_2C_2}^2)/(u(r_l)-u(\epsilon))$.
In a general situation, where $A_1\neq B_1$, the same analysis is also
valid around $x_1$.  The results are best expressed in terms of the
minimizing variables $C_i=B_i\sqrt{B_i^\dagger A_i}$ and the variables
$D_i$, defined by the relations
\begin{equation}
A_i = C_i D_i ,\ \  B_i = C_i D_i^\dagger,
\label{eq:defCD}
\end{equation}
which yields $D_i=\sqrt{B_i^\dagger A_i}$. For the minimizing $C_i$'s we have
\begin{equation}
s_{A_iC_i}^2 + s_{B_iC_i}^2=
\frac{s_{A_i B_i}^2}{2}=2\mathrm{Arccos}(\half\mathrm{Tr}(D_i^2)).
\label{dist}
\end{equation}
Now, the behaviors (at $r_l$) around $x_1$ and $x_2$ can be matched by
means of a time dependent ansatz whose minimization will complete the
construction of the saddle point and will represent a slowly rotating
soliton from $C_2$ to $C_1$.  Again, a better approximation is obtained
by considering free propagation over $\mathrm{SU(2)}$ from $C_2$ to
$C_1$ (whose expression can be found in~\cite{S81}); including the
factors generated in the unlocalization region, we obtain the following
expression for $\tilde{G}$
\begin{eqnarray}
\tilde{G}(x_1, A_1, B_1;x_2, A_2, B_2) &=& 
g_1(\mathrm{Tr} D_1)
g_2(\mathrm{Tr} D_2)
\sum_{\bar{I}} D^{\bar{I}}_{km}(C_1) 
               D^{\bar{I}}_{mk}(C_2^\dagger)\nonumber\\
&& 
\mathrm{e}^{-\frac{\bar{I}(\bar{I}+1)}{2\Im}|x_2-x_1|}
\frac{\mathrm{e}^{-M_{\mathrm{cl}}|x_2-x_1|}}{\left(\frac{2\pi}{M} 
|x_2-x_1|\right)^{3/2}},
\label{eq:Gexp}
\end{eqnarray}
where we also include the factor $\exp{-M_{\mathrm{cl}}|x_2-x_1|}$ coming from
the classical action, and the factor proportional to $|x_2-x_1|^{-3/2}$
resulting from the integration over the translational zero-modes. 
The function $g_i$ only depends on $\mathrm{Tr}(D_i)$, since for an 
$\mathrm{SU(2)}$ matrix $ \mathrm{Tr}(D^2)=(\mathrm{Tr}(D))^2-2$, and 
contains the gaussian factor 
$\exp(-(s_{A_iC_i}^2 + s_{B_iC_i}^2)/(u(r_l)-u(\epsilon))$ (cf. eq. (\ref{dist})). 

Finally, to obtain the propagator $G$, we note that $A\rightarrow -A$,
$B\rightarrow -B$ is equivalent to $C\rightarrow -C$, $D \rightarrow
D$, and $D^{\bar{I}}(-C)=(-)^{2\bar{I}} D^{\bar{I}}(C)$; therefore, if
we define $G$ in eq. (\ref{eq:GGtilde}) using the plus (minus) sign the
only surviving terms are those with integer (half-integer) $\bar{I}$.
We will show below that $\bar{I}$ is the spin and isospin, thus
choosing the plus (minus) sign will lead to integer (half-integer) spin
(equal to isospin) particles.

Note that $\tilde{G}$ in equation (\ref{eq:Gexp}) enjoys the two
following properties: (1) $\tilde{G}$ satisfies the parity requirement
(\ref{eq:parity}) (invariance under $A_i\leftrightarrow B_i$) since
this is equivalent to $D_i\leftrightarrow D_i^\dagger$ and for an
$\mathrm{SU(2)}$ matrix $\mathrm{Tr}(D)=\mathrm{Tr}(D^\dagger)$, (2)
$\tilde{G}$ is invariant under the new symmetry $D_i\rightarrow
R_iD_iR_i^\dagger$.

Now we will show that the propagator of eq. (\ref{eq:Gexp})
propagates spin equal to isospin particles.

To find the propagator of the fields $\psi^{j_1j_2}$ we use eq. (\ref{eq:psi})
to obtain:
\begin{eqnarray}
\lefteqn{\langle0|T\{\bar{\psi}^{j_3j_4I'}_{\sigma_3\sigma_4m'_I}
\psi^{j_1j_2I}_{\sigma_1\sigma_2m_I}\}|0\rangle =
Z_{j_1j_2I}^{-1} Z_{j_3j_4I'}^{-1}
\sum_{\bar{I}}\left(\begin{array}{lll}j_3&j_4&I'\\
\sigma''_3&\sigma''_4&m'_I\end{array}\right)
\left(\begin{array}{lll}j_1&j_2&I\\
\sigma''_1&\sigma''_2&m_I\end{array}\right)} \nonumber\\
&&
C_{\sigma'_1\sigma_1}
C_{\sigma'_2\sigma_2}
C_{\sigma'_3\sigma_3}
C_{\sigma'_4\sigma_4}
\int_{A_{1,2}B_{1,2}} 
D^{j_1}_{\sigma''_1\sigma'_1}(A_1)
D^{j_2}_{\sigma''_2\sigma'_2}(B_1)
D^{j_3}_{\sigma''_3\sigma'_3}(A_2)
D^{j_4}_{\sigma''_4\sigma'_4}(B_2) \nonumber\\
&&
g_1(\mathrm{Tr} D_1) g_2(\mathrm{Tr} D_2) 
D^{\bar{I}}_{km}(C_1) D^{\bar{I}}_{mk}(C_2^\dagger)
\frac{\mathrm{e}^{-M_{\bar{I}}|x_2-x_1|}}
{\left(\frac{2\pi}{M} |x_2-x_1|\right)^{3/2}},
\end{eqnarray}
where we introduced 
$M_{\bar{I}} = M_{\mathrm{cl}} + \bar{I}(\bar{I}+1) / (2\Im)$.
Let us first perform the integral over $A_1$, $B_1$. It is convenient
to write everything in terms of $C_1$ and $D_1$, the factors involved are:
\begin{eqnarray}
&& \left(\begin{array}{lll}j_1&j_2&I\\
\sigma''_1&\sigma''_2&m_I\end{array}\right)
C_{\sigma'_1\sigma_1} C_{\sigma'_2\sigma_2} \times\nonumber\\
&& \times \int_{A_1B_1} 
D^{j_1}_{\sigma''_1\sigma'_1}(C_1D_1)
D^{j_2}_{\sigma''_2\sigma'_2}(C_1D_1^\dagger)
g_1(\mathrm{Tr} D_1)
D^{\bar{I}}_{km}(C_1).
\end{eqnarray}
To change variables we must evaluate a jacobian which can be
calculated parametrizing the matrices as 
$A=a_0+\mathrm{i}\vec{a}.\vec{\tau}$ with $(a_0^2+\vec{a}^2=1)$ and 
similar expressions for $B$, $C$ and $D$. The invariant measure is written as
\begin{equation}
\d A=\frac{\d a_1 \d a_2 \d a_3}{a_0},
\end{equation}
The jacobian is the determinant of a $6\times 6$ matrix which upon
evaluation gives:
\begin{equation}
\int \d A \d B = \int \d C \d D \ 2 (\mathrm{Tr} D)^2.
\end{equation}
We absorb this factor in $g_1$ defining $\bar{g}_1=2 g_1 (\mathrm{Tr}D_1)^2$. 
Since $\bar{g}_1$ is a class function of $D_1$, that is $\bar{g}_1(SD_1S^\dagger)=\bar{g}_1(D_1)$, it can be expanded according to
\begin{equation}
\bar{g}_1(D_1) = \sum_{l_1} \bar{g}_{1l_1} D^{l_1}_{\bar{m}\bar{m}}(D_1),
\end{equation}
We are now ready to evaluate the integrals since the integral of the product 
of three $D$ matrices is given by the $3j$ symbols~\cite{E60}. The result is
\begin{eqnarray}
&&
\sum_{l_1} \bar{g}_{1l_1} 
C_{\sigma'_1\sigma_1}  C_{\sigma'''_2\sigma'_2} \nonumber\\
&&
\left(\begin{array}{lll}j_1&j_2&I\\\sigma''_1&\sigma''_2&m_I\end{array}\right)
\left(\begin{array}{lll}j_1&j_2&\bar{I}\\\sigma''_1&\sigma''_2&k\end{array}\right)
\left(\begin{array}{lll}j_1&j_2&\bar{I}\\\sigma'''_1&\sigma'''_2&m\end{array}\right)
\left(\begin{array}{lll}j_1&j_2&l_1\\\sigma'''_1&\sigma_2&\bar{m}\end{array}\right)
\left(\begin{array}{lll}j_1&j_2&l_1\\\sigma'_1&\sigma'_2&\bar{m}\end{array}\right).
\end{eqnarray} 
The two first $3j$ symbols give $\delta _{I\bar{I}} \delta_{m_Ik} /(2I+1)$, the 
three remaining symbols can be simplified by means of the $6j$ symbol to give
\begin{equation}
\sum_{l_1} \frac{\bar{g}_{1l_1}}{2I+1}  
C_{\sigma_1\sigma'_1}  C_{\sigma_2\sigma'_2} \delta_{km_I}
\left\{\begin{array}{lll}j_1&j_2&I\\j_1&j_2&l_1\end{array}\right\}
\left(\begin{array}{lll}j_1&j_2&I\\\sigma'_1&\sigma'_2&m\end{array}\right).
\end{equation}
The integral over $A_2$, $B_2$ can be performed in a similar way. Putting all 
the pieces together we obtain:
\begin{eqnarray}
\lefteqn{\langle0|T\{\bar{\psi}^{j_3j_4I'}_{\sigma_3\sigma_4m'_I}
\psi^{j_1j_2I}_{\sigma_1\sigma_2m_I}\}|0\rangle = 
 Z^{-1}_{j_1j_2I} \bar{Z}^{-1}_{j_3j_4I'} 
\frac{C_{m'_Im_I} \delta _{II'}}{(2I+1)^2}
\frac{\mathrm{e}^{-M_I|x_2-x_1|}}{\left(\frac{2\pi}{M} 
|x_2-x_1|\right)^{3/2}} }\nonumber\\
&&
\left(\begin{array}{lll}j_1&j_2&I\\\sigma_1&\sigma_2&m\end{array}\right)
\left(\begin{array}{lll}j_3&j_4&I\\\sigma_3&\sigma_4&m'\end{array}\right)
C_{m'm}  
\sum_{l_1l_2} \bar{g}_{1l_1}\bar{g}_{2l_2}
\left\{\begin{array}{lll}j_1&j_2&I\\j_1&j_2&l_1\end{array}\right\}
\left\{\begin{array}{lll}j_3&j_4&I\\j_3&j_4&l_2\end{array}\right\}. \nonumber\\
\end{eqnarray}
Choosing in this equation
\begin{equation}
Z_{j_1j_2I} = \sum_{l_1} \frac{\bar{g}_{1l_1}}{2I+1} 
\left\{\begin{array}{lll}j_1&j_2&I\\j_1&j_2&l_1\end{array}\right\},
\label{eq:Zdef}
\end{equation}
and the same for $Z_{j_3j_4I'}$ (replacing $\bar{g}_{1l_1}$ by
$\bar{g}_{2l_2}$) we arrive at 
\begin{equation}
\left(\begin{array}{lll}j_1&j_2&I\\\sigma_1&\sigma_2&m\end{array}\right)
\left(\begin{array}{lll}j_3&j_4&I\\\sigma_3&\sigma_4&m'\end{array}\right)
C_{m'm}  C_{m'_Im_I}
\delta _{II'} 
\frac{\mathrm{e}^{-M_I |x_2-x_1|}}{\left(\frac{2\pi}{M} 
|x_2-x_1|\right)^{3/2}}.
\end{equation}
Finally, from eq. (\ref{eq:prop_etab}) we can read the value of $\bar{\eta}$
and then extend the propagator from $X_{12}=1$ to an arbitrary $X_{12}$,
obtaining
\begin{eqnarray}
\langle0|T\{\bar{\psi}^{j_3j_4I'}_{\sigma_3\sigma_4m'_I}
\psi^{j_1j_2I}_{\sigma_1\sigma_2m_I}\}|0\rangle &=& 
C_{m'm}  C_{m'_Im_I} \delta _{II'}
{\cal D}^{j_1}_{\sigma_1 \sigma'_1}(X_{12})
{\cal D}^{j_4}_{\sigma_4 \sigma'_4}(X_{12})  \nonumber \\
&&
\left(\begin{array}{lll}j_1&j_2&I\\\sigma'_1&\sigma_2&m\end{array}\right)
\left(\begin{array}{lll}j_3&j_4&I\\\sigma_3&\sigma'_4&m'\end{array}\right)
\frac{\mathrm{e}^{-M_I |x_2-x_1|}}{\left(\frac{2\pi}{M} 
|x_2-x_1|\right)^{3/2}}.  
\label{CCd} 
\end{eqnarray}
This expression coincides with the free field propagator we evaluate in
appendix~\ref{apA}, namely, formula (\ref{eq:Wprop}), where the index
$I$ in (\ref{CCd}) replaces the index $j$ in that formula. The index
$j$ in (\ref{eq:Wprop}) is the spin of the one particle states we use
in the appendix to construct the free fields, on the other hand, $I$ is
the isospin of our fields (cf. eq. (\ref{eq:LABpsi})), implying that
the propagator $\tilde{G}$ describes a tower of spin equal to isospin
particles (with mass $M_I$). Furthermore, the Green functions obtained
in the appendix correspond to free fields that for integer
(half-integer) $j$ commute (anticommute) at space-like separations,
that is, the fields we define by means of the Green functions
(\ref{Greenfield}) and eq. (\ref{eq:psi}) obey the spin-statistics
theorem. For example, one can calculate from eq. (\ref{CCd}) the 
propagator in the $(1/2,0)\oplus(0,1/2)$ representation which coincides,
at large distances, with the usual Dirac propagator 
for a spin $1/2$ particle. 

\section{S-matrix definition} \label{sec:Yc}

In order to compute scattering amplitudes one should be able to extract
the S-matrix from the Green functions previously defined.  As a first
step we apply definition (\ref{eq:psi}) to obtain Green functions for
the fields $\psi^{j_1j_2}$ and $\bar{\psi}^{j_1j_2}$ (remember that
$\psi$ and $\bar{\psi}$ refer to singularities of topological number
$+1$ and $-1$ respectively). Then, the Green functions are analytically
continued to Minkowski space-time by the replacement $x^4\rightarrow
-\mathrm{i}x^0$. The large distance propagator, obtained in the
previous section, allows us to identify the asymptotic states to be
scattered, in the non-trivial topological sector, as composed by spin
equal to isospin particles.  Now, all we need is the corresponding LSZ
reduction formula, which is obtained in the appendix and reads
\begin{eqnarray}
\lefteqn{_{\mathrm{out}}
\langle p_1\bar{\sigma}_1\bar{s}_1,
\cdots,p_n\bar{\sigma}_n\bar{s}_n
 |q_1\sigma_1s_1,\cdots,q_l\sigma_ls_l
\rangle_{\mathrm{in}} =
\mbox{disc. parts} + (\mathrm{i}Z^{-1/2})^{n+l}} \nonumber\\
&& 
\int\d^4y_1\cdots\d^4x_l 
\bar{u}^{j_1j'_1I_1}_{m_1m'_1}(p_1,\bar{\sigma}_1)\cdots 
u^{j_lj'_lI_l}_{m_lm'_l}(q_l,\sigma_l) 
\exp\left(\mathrm{i}\sum_1^np_ky_k-\mathrm{i}\sum_1^lq_rx_r\right) \nonumber\\
&&
(\Box_{y_1} + m_1^2)\cdots(\Box_{x_l} + m_l^2)
\langle 0|T\bar{\psi}^{j_1j'_1I_1}_{m_1m'_1\bar{s}_1}(y_1)
\cdots\psi^{j_lj'_lI_l}_{m_lm'_ls_l}(x_l)
|0\rangle, \label{eq:LSZred}
\end{eqnarray}
where $\sigma_i$ stands for the spin projection $z$ in the rest frame
and $s$ is the isospin $3^{rd}$ projection.  The wave functions $u$,
$\bar{u}$ are defined in the appendix; when considering the scattering
of antiparticles, $u,\bar{u}$ must be replaced by $v,\bar{v}$. The Green 
function in eq. (\ref{eq:LSZred}) is obtained from (\ref{mixed}) and 
(\ref{eq:psi}). An intuitive understanding of this Green function 
can be achieved using the line defects picture of ref.~\cite{FM90}. There,
the leading contribution to the path integral (\ref{mixed}) is identified 
with configurations described by open line defects emerging from
$x_1,...,x_n$ plus a gas of closed defect tubes. If one brings, e.g., $x_1$ 
far from the other points, the sum over the open line defects associated 
with $x_1$ will give a skyrmion free propagator, from $x_1$ towards the 
interaction region. This propagator has the pole (in $p^2$) required
for a non-vanishing scattering amplitude and also contains all information
about the boundary condition around $x_1$.  

As it is noticed in the appendix, the different fields that can be
built from the same particle states are not independent and one should
choose only one $\psi^{j_1j_2I}$ to represent each kind of particle.
For example the $\Delta$ $(j=3/2)$ can be represented by a $(3/2,0)$
field or by a $(1,1/2)$, as in the Rarita-Schwinger representation.
Choosing one or the other amounts to a field redefinition and hence the
S-matrix should be independent of this choice. This assertion can be
checked, within the saddle point approximation, taking into account
that all information about the boundary conditions is contained in the
free propagators (\ref{CCd}) attached to the points $x_i$ (as described
in the previous paragraph). The explicit dependence on $(j_1,j_2)$ is
cancelled by the choice of normalization factor $Z_{j_1j_2I}$
(\ref{eq:Zdef}).
 
It is interesting to note that, for a $(1/2,0)$ field, the reduction
formula we give contains an operator $(\Box+m^2)$, instead of the usual
$(\mathrm{i}\FMslash{\partial} +m)$. This comes about because we are
using a two-component spinor, see~\cite{FG58} for a discussion on this 
point. For a two-component spinor we need $\psi^{1/2\,\,0}$ and $\partial_t\psi^{1/2\,\,0}$ to recover the
operators $a$ and $b^\dagger$. In the case of spin $1/2$ fermions, the
usual alternative approach is considering two fields, namely
$\psi^{1/2\,\,0}$ and $\psi^{0\,\,1/2}$ (or in general $\psi^{j_1j_2}$
and $\psi^{j_2j_1}$). Using these two fields ($4$-component spinors)
one can obtain $a$ and $b^\dagger$ with no need of the field time
derivative.  Then, in that case, the LSZ reduction formula displays a
first order operator. The same can be done for higher spin fields but
the operator appearing in the LSZ formula will be of higher order and
more cumbersome to use than the operator $(\Box+m^2)$ associated with a
single field representation.

The Green functions we defined in eq. (\ref{mixed}) imply
that our pion field is not just the fluctuating part but the sum of the
classical configuration plus the fluctuating part. In fact, this is the
natural thing to do, the separation between classical and fluctuating
parts is just a consequence of the saddle point approximation, which
may not always be adequate.

Precisely, the suggestion that the pion field be identified with the
classical soliton plus fluctuations, instead of identifying it just
with the fluctuations, led to the solution to the Yukawa problem (see
refs.~\cite{DHM94}, \cite{DPP} and \cite{HSU}).

The Yukawa coupling is obtained from the vertex
\begin{equation}
\langle 0| T\{ \psi^{j_1,j_2,I}_{\sigma_1,\sigma_2,m_I}(x_1)
\bar{\psi}^{j_4,j_3,I'}_{\sigma_4,\sigma_3,m'_I}(x_2) \phi_a(y_1)\}|0 \rangle,
\label{eq:vertex}
\end{equation}
At large distances ($|x_1-x_2|$ large) the relevant configurations are
described in terms of open line defects which can be interpreted as
skyrmion world lines.  Then, in this case, the saddle point calculation
is similar to that of reference~\cite{DHM94} where the correct Yukawa
coupling was obtained resorting to a rotationally improved ansatz. In
our formalism, the calculation would be anologous to the evaluation of
the propagator performed in section~\ref{sec:Sp}.

At short distances ($|x_1-x_2|$ small) the saddle point no longer
resembles a static solution in the interval between creation and
destruction of the soliton, so one should expect differences with the
approach of~\cite{DHM94}. However we will not pursue this point here
since at short distances the Skyrme lagrangian is not adequate.
However, this discussion may be relevant in connection with other
solitonic models.

\section{Conclusions} \label{sec:C}

In ref.~\cite{FM90}, Fr\"ohlich and Marchetti proposed a quantum field
theory for skyrmions, defining Euclidean Green functions in terms of
path integrals over singular fields.

In the present paper we show that by fixing the boundary conditions
around the singularities it is possible to study the physical content
of the Skyrme model and their dynamical implications in a manifestly
covariant way.  In particular, we obtain the covariant baryon
propagator. This treatment may also be relevant when studying other
solitonic models as for example monopoles.

Our method is based on defining Green functions subject to a minimal
set of boundary conditions that close under four dimensional rotations 
and isospin transformations. The fields $\psi_{A,B}$ defined in this way are
labeled by two $\mathrm{SU(2)}$ matrices and carry an infinite
(reducible) representation of the isospin group and four dimensional rotations.
 Equivalently, we can define a set of fields $\psi^{j_1 j_2 I}$
transforming under finite dimensional irreducible representations.
 Here, the spin of the field $\psi^{j_1 j_2 I}$ ranges, in principle,
from $|j_2-j_1|$ to $j_1+j_2$ ($I$ is the isospin); for these fields we
define the corresponding Green functions and derive the LSZ formulae to
compute the S-matrix, thus mapping a complete relativistic field
theory. The S-matrix constraints of locality and unitarity are expected to be 
satisfied. An heuristic argument could be given along the lines 
of ref.~\cite{FM87,FM90}, using the Osterwalder-Schrader reconstruction theorem.

The numerical calculation of the saddle point shows that near the
singularities we have a region where the chiral symmetry is unbroken,
while it is broken away from the singularities. These regions
correspond to the unlocalization region, where the topological density
is isotropically distributed in Euclidean space-time, and the
localization region, where the topological density gets concentrated,
signaling the presence of a soliton.

In order to compute the path integrals, we show that, in the
localization region, and when the skyrmion is created at rest, the
relevant paths can be characterized by (non-covariant) variables
$C=B\sqrt{B^{\dagger}A}$, obtained by minimizing the action generated
in the unlocalization region.

The covariant propagator is evaluated by using symmetry arguments and
integrating over zero modes in the localization region. The result
obtained shows a new symmetry under the transformation $D\rightarrow
RDR^{\dagger}$, where $D=\sqrt{B^{\dagger}A}$. This symmetry is seen to
be responsible, upon an explicit calculation, for the spin equal to
isospin skyrmion spectrum of the model.

Another consequence that shows up, upon fixing the boundary conditions,
is the introduction of additional topology, which leads to consider 
the extension of a map $U:S^3\times S^1\rightarrow \mathrm{SU(2)}$ 
to a map $U:S^3\times D_2 \rightarrow\mathrm{SU(2)}$ to study the 
different possible inequivalent quantizations of the 
$\mathrm{SU(2)}$-Skyrme model~\cite{Wi83}.  There
are two possibilities, one of them corresponding to the baryonic
spectrum composed by half-integer spin (equal to isospin) particles,
which are seen to satisfy the spin-statistics theorem.

Finally, we think that our method also provides a formal base for the 
work in ref.~\cite{DHM94} where a skyrmion field theory is advocated 
(together with an improved approximation scheme) to solve the Yukawa 
problem, that is, reproducing the pseudovector pion-baryon coupling from
skyrmion physics.

\ack

We are grateful to D. Mazzitelli, N.N. Scoccola and E.C. Marino for 
interesting discussions. Collaboration with J.P. Garrahan and C.L. Schat
during the first stage of this work is also acknowledged.

\appendix

\section{}
\label{apA}

In this appendix we summarize a method due to Weinberg~\cite{W64}
to define arbitrary spin field propagators in Minkowski
space.

We start by considering states containing one particle (with mass $m$) at rest, 
having spin $j$ and spin projection $\sigma$: $|(m,0);j\sigma\rangle$ and
then we define the states
\begin{equation}
|p;j\sigma> = \sqrt{\frac{m}{p_0}} U(L_p) |(m,0);j\sigma\rangle,
\label{eq:states}
\end{equation}
where $U(L_p)$ is a boost that maps $(m,0)$ into $p$, which is defined
by the $\mathrm{SL}(2,C)$ matrix
\begin{equation}
L_p = \exp(\theta \hat{p} . \vec{\tau}).
\end{equation}
where $\hat{p}$ is the unit vector $\vec{p}/|\vec{p}|$, and 
$\sinh(\theta) = |\vec{p}|/m$. Note that 
$L_p L_p = (p_0 + \vec{p}.\vec{\tau})/m$.
The Lorentz transformations act over the states (\ref{eq:states})
 according to:
\begin{eqnarray}
U_\Lambda |p;j\sigma\rangle &=& \sqrt{\frac{m}{p_0}} U(\Lambda L_p) 
 |(m,0);j\sigma\rangle =  \sqrt{\frac{m}{p_0}} U(L_{\Lambda p})
U(L_{\Lambda p}^{-1} \Lambda L_p) |(m,0);j\sigma\rangle \nonumber\\
&=&   \sqrt{\frac{(\Lambda p)_0}{p_0}} 
D^j_{\sigma'\sigma}(L_{\Lambda p}^{-1} \Lambda L_p)
 |\Lambda p;j\sigma'\rangle.
\label{eq:st}
\end{eqnarray}
The key point is that $L_{\Lambda p}^{-1} \Lambda L_p$ is 
a rotation and rotations act on the states $|(m,0);j\sigma\rangle$
by means of the $D^j$ matrices, as usual.

Particles in the state $|p;j\sigma\rangle$ are created from the vacuum by operators
$a^\dagger_{p,\sigma}$ which, in view of eq.  (\ref{eq:st}), satisfy:
\begin{eqnarray}
U_\Lambda a_{p,\sigma} U_\Lambda^\dagger &=&  \sqrt{\frac{(\Lambda p)_0}{p_0}} 
D^j_{\sigma\sigma'}(L_p^{-1}\Lambda^{-1}L_{\Lambda p}) a_{\Lambda p,\sigma'},
 \nonumber\\
U_\Lambda a^\dagger_{p,\sigma} U_\Lambda^\dagger &=& 
\sqrt{\frac{(\Lambda p)_0}{p_0}} 
(CD^j(L_p^{-1}\Lambda^{-1}L_{\Lambda p})C^{-1})_{\sigma\sigma'}
 a^\dagger_{\Lambda p,\sigma'}.
\end{eqnarray}
Using the operators $a,a^\dagger$, and the corresponding operators $b,b^\dagger$ 
for the antiparticles, fields in any finite dimensional
representation $(j_1,j_2)$ of the Lorentz group (with
$|j_2-j_1|<j<|j_1+j_2|$) can be constructed through the formulae
\begin{eqnarray}
\psi^{j_1j_2}_{m_1m_2} &=& \int \frac{d^3p}{(2\pi)^{3/2}}
\frac{1}{\sqrt{2\omega_p}} 
{\cal D}^{j_1}_{m_1m'_1}(L_p) \bar{{\cal D}}^{j_2}_{m_2m'_2}(L_p) \nonumber\\
&& C_{\sigma\sigma'} 
\left(\begin{array}{lll}j_1&j_2&j\\ m'_1&m'_2&\sigma'\end{array}\right)
\left[ a_{p,\sigma} \mathrm{e}^{-ipx}
+  (-)^{2j_2} C^{-1}_{\sigma\sigma''} b^\dagger_{p,\sigma''} 
\mathrm{e}^{ipx} \right], \nonumber\\
\bar{\psi}^{j_1j_2}_{m_1m_2} &=& C_{m_1'm_1} C_{m_2' m_2} 
\left( \psi^{j_1j_2}_{m'_1m'_2}\right) ^{\dagger}.
\label{eq:freef}
\end{eqnarray}
The sign $(-)^{2j_2}$ is needed for the $\psi$ and $\psi^\dagger$ to
commute (or anticommute) at space-like separations. At time-like
separations they do not commute since they are built with the same
operators $a,a^\dagger$. This means that the various relativistic fields
we can build from the same spin $j$ particle states are not
independent; we must choose one of them to represent these particles.
The field $\psi$ transforms as
\begin{equation}
\mathrm{U}_\Lambda \psi^{j_1j_2}_{m_1m_2}(x) \mathrm{U}_\Lambda^\dagger =
{\cal D}^{j_1}_{m_1m'_1}(\Lambda^{-1})
\bar{{\cal D}}^{j_2}_{m_2m'_2}(\Lambda^{-1}) \psi^{j_1j_2}_{m'_1m'_2}
(\Lambda x)
\end{equation}
while $\bar{\psi}$ obeys a similar relation but with $j_1,j_2$ interchanged,
as corresponds to the adjoint operator. 
Now we can readily evaluate the propagator
\begin{equation}
\langle 0|T\{ \psi(x)^{j_1j_2}_{m_1m_2} 
\bar{\psi}(y)^{j_3j_4}_{m_3m_4}
\} |0\rangle =
\left\{ \begin{array}{lll}
\langle 0|\psi(x)^{j_1j_2}_{m_1m_2}
\bar{\psi}(y)^{j_3j_4}_{m_3m_4}
|0\rangle&\mbox{if}&x_0>y_0 \\
(-)^{2j} \langle 0|\bar{\psi}(y)^{j_3j_4}_{m_3m_4}
\psi(x)^{j_1j_2}_{m_1m_2}
 |0\rangle
&\mbox{if}&y_0>x_0 \end{array} \right.
\end{equation}
giving 
\begin{eqnarray}
\int && \frac{d^3p}{(2\pi)^32\omega_p}  
{\cal D}^{j_1}_{m_1m'_1}(L_p) 
\bar{{\cal D}}^{j_2}_{m_2m'_2}(L_p) 
\bar{{\cal D}}^{j_3}_{m_4m'_4}(L_p) 
{\cal D}^{j_4}_{m_3m'_3}(L_p) 
\left(\begin{array}{lll}j_1&j_2&j\\ m'_1&m'_2&\sigma\end{array}\right)
C_{\sigma\sigma'} \nonumber\\
&&
\left(\begin{array}{lll}j_3&j_4&j\\ m'_3&m'_4&\sigma'\end{array}\right)
\left[\theta(x_0-y_0) \mathrm{e}^{-ip(x-y)} + (-)^{(2j_2+2j_4)}
      \theta(y_0-x_0) \mathrm{e}^{ ip(x-y)} \right]
\end{eqnarray}
For a boost $\bar{{\cal D}}^j(L_p)={\cal D}^j(L_p^{-1})$; then, 
using the property of the $3j$ symbols:
\begin{equation}
{\cal D}^{j_1}_{m_1m'_1}(L_p^{-1}) 
\left(\begin{array}{lll}j_1&j_2&j_3\\m'_1&m_2&m_3\end{array}\right)=
{\cal D}^{j_2}_{m_2m'_2}(L_p) 
{\cal D}^{j_3}_{m_3m'_3}(L_p) 
\left(\begin{array}{lll}j_1&j_2&j_3\\m_1&m'_2&m'_3\end{array}\right).
\end{equation}
This expression can be transformed into
\begin{eqnarray}
\int \frac{d^3p}{(2\pi)^32\omega_p} && 
{\cal D}^{j_1}_{m_1m'_1}(L_pL_p) 
{\cal D}^{j_4}_{m_4m'_4}(L_pL_p) 
C_{\sigma'\sigma}
\left(\begin{array}{lll}j_1&j_2&j\\ m_1&m'_2&\sigma'\end{array}\right)
\left(\begin{array}{lll}j_3&j_4&j\\ m_3&m'_4&\sigma\end{array}\right)\nonumber\\
&&\left[\theta(x_0-y_0) \mathrm{e}^{-ip(x-y)} +  (-)^{(2j_2+2j_4)}
      \theta(y_0-x_0) \mathrm{e}^{ ip(x-y)} \right].
\label{integralA}
\end{eqnarray}
Taking into account that 
\begin{equation}
{\cal D}^{1/2}(L_pL_p)=L_pL_p=\frac{1}{m}(p_0 + \vec{p}.\vec{\tau})
\end{equation}
and that the elements of the matrix ${\cal D}^j$ are linear combinations
of the elements of ${\cal D}^{1/2} \otimes \ldots \otimes {\cal D}^{1/2}$,
($2j$ times), it follows that ${\cal D}^j$ is a polynomial in $p_\mu$ of order
$2j$. In the integral (\ref{integralA}) we can ``take out'' the polynomials 
${\cal D}^j((p_0+\vec{p}.\vec{\tau})/m)$, if we replace $p_\mu$ by 
$-\mathrm{i}\partial_\mu$. So the final expression for the Minkowski propagator
is 
\begin{eqnarray}
&& \langle 0|T\{ \psi(x)^{j_1,j_2}_{m_1,m_2} 
\bar{\psi}(y)^{j_3,j_4}_{m_3,m_4}
\} |0\rangle =
C_{\sigma\sigma'} 
\left(\begin{array}{lll}j_1&j_2&j\\ m'_1&m_2&\sigma'\end{array}\right)
\left(\begin{array}{lll}j_3&j_4&j\\ m_3&m'_4&\sigma\end{array}\right)
\nonumber\\
&& {\cal D}^{j_1}_{m_1m_1'}
(-\frac{\mathrm{i}}{m}(\partial_0+\vec{\tau}.\vec{\partial}))
 {\cal D}^{j_4}_{m_4m_4'}
(-\frac{\mathrm{i}}{m}(\partial_0+\vec{\tau}.\vec{\partial}))
\Delta_F(x-y)  ,
\end{eqnarray}
where we introduced the scalar Feynman propagator $\Delta_F$. When extracting
the time derivatives we did not applied them to the $\theta$ functions
that define the time ordering. Thus, the above propagator is the so called
covariant propagator. The non covariant pieces should be automatically 
cancelled by non covariant vertices in the 
interaction Hamiltonian when using the canonical formalism (see~\cite{W64}). 
In our case, since the path integral used to define the propagator 
(eq. (\ref{eq:propG})) is Lorentz invariant, we also expect to obtain this 
covariant propagator. As the ${\cal D}^{j}(-\mathrm{i}\sigma^\mu \partial_mu)$
are polynomials in $\partial_\mu$, this expression is suitable
 for analytic continuation into Euclidean space. Defining the Euclidean 
propagator as $\Delta_E(x_4,\vec{x}) = \Delta_F(\mathrm{i}x_4,\vec{x})$
we obtain the analytic continuation
\begin{eqnarray}
 &&
C_{\sigma\sigma'}
\left(\begin{array}{lll}j_1&j_2&j\\ m'_1&m_2&\sigma'\end{array}\right)
\left(\begin{array}{lll}j_3&j_4&j\\ m_3&m'_4&\sigma\end{array}\right)
\nonumber\\
&&
 {\cal D}^{j_1}_{m_1m_1'}
(-\frac{1}{m} (\partial_0 + \mathrm{i} \vec{\tau}.\vec{\partial}))
 {\cal D}^{j_4}_{m_4m_4'}
(-\frac{1}{m} (\partial_0 + \mathrm{i} \vec{\tau}.\vec{\partial}))
\Delta_E(x-y).
\end{eqnarray}
For large distances $\Delta_E(x)$ behaves as
\begin{equation}
\Delta_E(x) \simeq_{|x|\rightarrow \infty} \frac{\mathrm{e}^{-m|x|}}
{\left(\frac{2\pi}{m} |x|\right)^{3/2}},
\end{equation}
and to leading order in $1/(m|x|)$ we can replace $\partial_\mu$ by
$-mx_\mu/x$, giving
\begin{equation}
{\cal D}^{j_1}_{m_1m_1'}(X)
{\cal D}^{j_4}_{m_4m_4'}(X)
C_{\sigma\sigma'} 
\left(\begin{array}{lll}j_1&j_2&j\\ m'_1&m_2&\sigma'\end{array}\right)
\left(\begin{array}{lll}j_3&j_4&j\\ m_3&m'_4&\sigma\end{array}\right)
 \frac{\mathrm{e}^{-m|x-y|}}
{\left(\frac{2\pi}{m} |x-y|\right)^{3/2}},
\label{eq:Wprop}
\end{equation}
where we introduced the matrix $X=x_4 + \mathrm{i} \vec{x}.\vec{\tau}$.

Another useful calculation that can be done with the free fields 
(\ref{eq:freef}) is the LSZ reduction formula. The procedure
is well-known~\cite{IZ80}. First 
we must express $a$, $a^\dagger$, $b$, $b^\dagger$ in terms of
the fields $\psi$. This is an easy task which gives, for example,
the $a^\dagger$ operator as:
\begin{eqnarray}
a^{\dagger(\mathrm{in})}_{p\sigma} &=& \frac{2j+1}{\sqrt{2\omega_p}} (-)^{2j} 
\left(\begin{array}{lll}j_1&j_2&j\\m_1&m_2&\sigma\end{array}\right)
\bar{{\cal D}}^{j_1}_{m_1m'_1}(L^{-1}_p) 
{\cal D}^{j_2}_{m_2m'_2}(L^{-1}_p) \nonumber\\
&&\int \frac{\d^3x}{(2\pi)^{3/2}}
\mathrm{e}^{\mathrm{i}\vec{p}\vec{x}}\mathrm{e}^{-\mathrm{i}\omega_pt}
\left(\omega_p \bar{\psi}^{(\mathrm{in})}_{m'_1m'_2} 
- \mathrm{i}\partial_t\bar{\psi}^{(\mathrm{in})}_{m'_1m'_2}\right).
\end{eqnarray}
The LSZ reduction proceeds now by using the asymptotic condition for
the interacting field
\begin{equation}
\langle f| \psi(t)|i\rangle \stackrel{t\rightarrow-\infty}{\longrightarrow}  
Z^{-1/2} \langle f| \psi^{(\mathrm{in})}|i\rangle, 
\end{equation}
to replace $\psi^{(\mathrm{in})}$ by $\psi$ in the formula for 
$a^\dagger_{p,\sigma}$. 
The renormalization factor $Z$ will absorb the
factor $\exp(-C_\epsilon)$ introduced in eq. (\ref{eq:propG}).
Then, using the relation
\begin{equation}
\lim_{t\rightarrow-\infty} f(t) = -\int_{-\infty}^{+\infty}\partial_t f(t) \d t
+ \lim_{t\rightarrow+\infty} f(t)
\end{equation}
we arrive at the reduction formula. The result for $a^\dagger$ 
as well as for the other operators is 
\begin{eqnarray}
\langle\mathrm{out}|a_{p\sigma}^{\dagger(\mathrm{in})} |\mathrm{in}\rangle &=&
\langle\mathrm{out}|a_{p\sigma}^{\dagger(\mathrm{out})} |\mathrm{in}\rangle +
\mathrm{i} Z^{1/2} 
\frac{2j+1}{\sqrt{2\omega_p}} (-)^{2j} 
\left(\begin{array}{lll}j_1&j_2&j\\m_1&m_2&\sigma\end{array}\right) 
\bar{{\cal D}}^{j_1}_{m_1m'_1}(L^{-1}_p) \nonumber\\
&&
{\cal D}^{j_2}_{m_2m'_2}(L^{-1}_p) 
\int \frac{\d^4x}{(2\pi)^{3/2}}
\mathrm{e}^{-\mathrm{i}px} (\Box +m^2) 
\langle\mathrm{out}|\bar{\psi}^{j_1j_2}_{m'_1m'_2} |\mathrm{in}\rangle \nonumber \\
&=& 
\mbox{disc.} + \mathrm{i} Z^{1/2}\int \d^4x\mathrm{e}^{-\mathrm{i}px} 
\bar{u}^{j_1j_2}_{m_1m_2}(p\sigma)(\Box +m^2) 
\langle\mathrm{out}|\bar{\psi}^{j_1j_2}_{m_1m_2} |\mathrm{in}\rangle, \nonumber\\
%---------
\langle\mathrm{out}|b_{p\sigma}^{\dagger(\mathrm{in})} |\mathrm{in}\rangle &=&
\langle\mathrm{out}|b_{p\sigma}^{\dagger(\mathrm{out})} |\mathrm{in}\rangle +
\mathrm{i} Z^{1/2} 
\frac{2j+1}{\sqrt{2\omega_p}} (-)^{2j_1} 
\left(\begin{array}{lll}j_1&j_2&j\\m_1&m_2&\sigma\end{array}\right)
{\cal D}^{j_1}_{m_1m'_1}(L^{-1}_p) \nonumber\\
&&
\bar{{\cal D}}^{j_2}_{m_2m'_2}(L^{-1}_p)
\int \frac{\d^4x}{(2\pi)^{3/2}}
\mathrm{e}^{-\mathrm{i}px} (\Box +m^2) 
\langle\mathrm{out}|\psi^{j_1j_2}_{m'_1m'_2} |\mathrm{in}\rangle \nonumber\\
&=&
\mbox{disc.} + \mathrm{i} Z^{1/2} \int \d^4x\mathrm{e}^{-\mathrm{i}px} 
v^{j_1j_2}_{m_1m_2}(p\sigma)(\Box +m^2) 
\langle\mathrm{out}|\psi^{j_1j_2}_{m_1m_2} |\mathrm{in}\rangle, \nonumber\\
%---------------
\langle\mathrm{out}|a_{p\sigma}^{(\mathrm{out})} |\mathrm{in}\rangle &=&
\langle\mathrm{out}|a_{p\sigma}^{(\mathrm{in})} |\mathrm{in}\rangle +
\mathrm{i} Z^{1/2} 
\frac{2j+1}{\sqrt{2\omega_p}} (-)^{2j} C_{\sigma\sigma'}
\left(\begin{array}{lll}j_1&j_2&j\\m_1&m_2&\sigma'\end{array}\right) 
{\cal D}^{j_1}_{m_1m'_1}(L^{-1}_p) \nonumber\\
&&
\bar{{\cal D}}^{j_2}_{m_2m'_2}(L^{-1}_p) 
\int \frac{\d^4x}{(2\pi)^{3/2}}
\mathrm{e}^{\mathrm{i}px} (\Box +m^2) 
\langle\mathrm{out}|\psi^{j_1j_2}_{m'_1m'_2} |\mathrm{in}\rangle \nonumber\\
&=& 
\mbox{disc.} + \mathrm{i} Z^{1/2} \int \d^4x\mathrm{e}^{\mathrm{i}px} 
u^{j_1j_2}_{m_1m_2}(p\sigma)(\Box +m^2) 
\langle\mathrm{out}|\psi^{j_1j_2}_{m_1m_2} |\mathrm{in}\rangle, \nonumber\\
%---------------
\langle\mathrm{out}|b_{p\sigma}^{(\mathrm{out})} |\mathrm{in}\rangle &=&
\langle\mathrm{out}|b_{p\sigma}^{(\mathrm{in})} |\mathrm{in}\rangle +
\mathrm{i} Z^{1/2} 
\frac{2j+1}{\sqrt{2\omega_p}} (-)^{2j_1}C_{\sigma\sigma'} 
\left(\begin{array}{lll}j_1&j_2&j\\m_1&m_2&\sigma'\end{array}\right)
\bar{{\cal D}}^{j_1}_{m_1m'_1}(L^{-1}_p) \nonumber\\
&&
{\cal D}^{j_2}_{m_2m'_2}(L^{-1}_p)
\int \frac{\d^4x}{(2\pi)^{3/2}}
\mathrm{e}^{\mathrm{i}px} (\Box +m^2) 
\langle\mathrm{out}|\bar{\psi}^{j_1j_2}_{m'_1m'_2} |\mathrm{in}\rangle \nonumber\\
&=& 
\mbox{disc.} + \mathrm{i} Z^{1/2} \int \d^4x\mathrm{e}^{\mathrm{i}px} 
\bar{v}^{j_1j_2}_{m_1m_2}(p\sigma)(\Box +m^2) 
\langle\mathrm{out}|\bar{\psi}^{j_1j_2}_{m_1m_2} |\mathrm{in}\rangle, 
\end{eqnarray}
where for brevity we defined $u$, $\bar{u}$, $v$, $\bar{v}$ which can be
interpreted as wave functions of incoming and outgoing particles.
These formulae allow us to extract the scattering amplitudes from 
the Green functions we defined.

\end{document}